# Coherent Interactions of Free Electrons and Matter: Toward Tunable Compact X-ray Sources

Amnon Balanov[1], Alexey Gorlach[1], Vladimir Baryshevsky[2], Ilya Feranchuk[3], Hideo Nitta[4], Yasushi Hayakawa[5], Alexander Shchagin[6,7], Yuichi Takabayashi[8], Yaron Danon[9], Liang Jie Wong [10], and Ido Kaminer[1]*

[1]*Department of Electrical and Computer Engineering, Technion – Israel Institute of Technology, Haifa 3200003, Israel*

[2]*Research Institute for Nuclear Problems of Belarusian State University Bobruiskaya Str. 11, 220030 Minsk, Belarus*

[3]*Physics Department, Belarusian State University, Nezavisimosti Ave, 4, 220030, Minsk, Belarus*

[4]*Department of Physics, Tokyo Gakugei University, Koganei', Tokyo 184, Japan*

[5]*Laboratory for Electron Beam Research and Application, Nihon University, Narashinodai 7-24-1, Funabashi 274-8501, Japan*

[6]*Deutsches Elektronen-Synchrotron DESY, Notkestr. 85, 22607 Hamburg, Germany*

[7]*Kharkiv Institute of Physics and Technology, Academicheskaya 1, Kharkiv 61108, Ukraine*

[8]*SAGA Light Source, 8-7 Yayoigaoka, Tosu, Saga 841-0005, Japan*

[9]*Department of Mechanical, Aerospace, and Nuclear Engineering, Rensselaer Polytechnic Institute, Troy, New York 12180, USA*

[10]*School of Electrical and Electronic Engineering, Nanyang Technological University, 50 Nanyang Ave, Singapore 639798, Singapore*

*corresponding author: kaminer@technion.ac.il

**Compact laboratory-scale X-ray sources still rely on the same fundamental principles as in the first X-ray tubes developed more than a century ago. In recent years, significant research and development have focused on large-scale X-ray sources such as synchrotrons and free-electron lasers, leading to the generation of high-brightness coherent X-rays. However, the large size and high costs of such sources prevent their widespread use. The quest for a compact and coherent X-ray source has long been a critical objective in modern physics, gaining further importance in recent years for industrial applications and fundamental scientific research. Here, we review the physical mechanisms governing compact coherent X-ray generation. Of current interest are coherent periodic interactions of free electrons in crystalline materials, creating hard X-rays via a mechanism known as parametric X-ray radiation (PXR). Over the past decade, X-ray sources leveraging this mechanism have demonstrated state-of-the-art tunability, directionality, and broad spatial coherence, enabling X-ray phase-contrast imaging on a compact scale. The coming years are expected to show substantial miniaturization of compact X-ray sources, facilitated by progress in electron beam technologies. This review compares the most promising mechanisms used for hard-X-ray generation, contrasting parametric X-ray radiation with inverse Compton scattering and characteristic radiation from a liquid-jet anode. We cover the most recent advancements, including the development of new materials, innovative geometrical designs, and specialized optimization techniques, aiming toward X-ray flux levels suitable for medical imaging and X-ray spectroscopy in compact scales.**

# Contents

# 1. Introduction

Since the discovery of X-ray radiation by Wilhelm Röntgen in 1895 [1], X-rays have revolutionized modern science and played a central role in many commercial and scientific applications. X-rays had a major impact on a wide range of fields, including medical imaging, biology, material science, environmental and earth science, astrophysics, homeland security, and industrial inspection. Indeed, X-ray science is responsible for numerous Nobel prizes in physics, chemistry, and medicine. A list of just part of the physics awards includes the 1901 award to Wilhelm Röntgen for the discovery of X-ray radiation, the 1914 award to Max von Laue for the discovery of X-ray diffraction by crystals, the 1915 award to William and Lawrence Bragg for the development of X-ray crystallography, the 1917 award to Charles Glover Barkla for the discovery of characteristic X-ray of elements, and the 1924 award to Karl Manne Georg Siegbahn for the discovery of the X-ray spectroscopy. It is quite remarkable that most of these discoveries were made using the relatively simple sources of hard X-rays based on Röntgen's compact X-ray tube. Notable exceptions include experiments using X-rays from radioactive elements, or measurements of X-rays arriving from astronomical phenomena in deep space.

Over the past decades, X-ray science has evolved along two distinct paths separated by the types and scales of X-ray sources: large-scale X-ray facilities vs compact X-ray sources. *Synchrotrons and free-electron laser (FEL) facilities*, which represent the pinnacle of X-ray technology, providing coherent, tunable hard X-rays with high flux and exceptional beam quality [2,3]. However, these facilities have significant drawbacks, including immense space requirements, high energy consumption, extensive safety measures, and limited accessibility due to their scale and cost. In contrast, *compact X-ray sources have primarily relied on X-ray tubes*, which are widely available and relatively inexpensive but emit isotropic and broadband radiation, lacking the energy tunability and coherence needed for many advanced applications. This contrast has driven ongoing research efforts toward novel compact mechanisms of X-ray generation that achieve the coherence and tunability of large-scale facilities without their associated drawbacks.

*Overview of coherent X-ray generation in compact scales*

The leading mechanisms for compact X-ray sources include high-harmonic generation (HHG) [4–15] laser-plasma accelerators (LPA) [16–26], inverse Compton scattering (ICS) [27–32], radioactive elements, and mechanisms based on the coherent interaction between free electrons and matter. The latter include Cherenkov radiation, Smith-Purcell radiation, channeling radiation, coherent Bremsstrahlung, transition radiation, and parametric X-ray radiation (PXR) [33,34]. Unlike the conventional X-ray tube, which is broadband and isotropic, these compact mechanisms offer varying degrees of coherence, flux, and energy tunability. They also differ in operational complexities, such as precision, shielding requirements, and other practical considerations.

The field of compact X-ray science is constantly evolving, with recent years witnessing the emergence of new concepts and mechanisms for X-ray generation. Novel mechanisms include the use of free-electron interactions with graphene surface plasmons, magnetic nanoundulators, metasurfaces, and metamaterials [35–41]. Improved designs in the free-electron sources themselves also rely on advanced materials such as carbon nanotubes [42]. Advances in high-intensity pulsed lasers inspired proposals of laser-undulators of electrons, both in vacuum and in specially tailored photonic waveguides [43–51]. In the realm of quantum electrodynamics and quantum optics, innovative theoretical proposals include the manipulation of vacuum fluctuations and the engineering of electron wavefunctions to enhance X-ray generation [52–57]. Parallel efforts in recent years showcase, pioneering experimental studies of X-ray generation in compact scales that now rely on the precision of electron microscopes, exploring novel structures such as van der Waals materials as electron undulators [58–60]. Contemporary experiments in the optical domain inspire new concepts for X-ray generation, as with Smith-Purcell lenses [61] and radiation enhancement based on photonic crystal flatbands [62].

*The need for compact sources of hard X-rays*

Many of these ongoing efforts are focused on the generation of *hard* X-rays. A directional hard X-ray source with a narrow spectral linewidth would be highly advantageous for many applications, allowing for a significant reduction in radiation dose [63]. For example, mammographic examinations using nearly mono-energetic X-rays could reduce the radiation dose by a factor of ten to fifteen compared to conventional X-ray systems [64]. Similar dose reductions are estimated for angiography and other radiographic studies [64].

There are currently three leading mechanisms for compact sources of *hard* X-rays with sufficient coherence and flux for imaging applications: parametric X-ray radiation (PXR), inverse Compton scattering (ICS), and characteristic radiation from liquid jets [65–67]. A detailed comparison is provided below. These mechanisms are analyzed and being developed for practical applications in medicine, homeland security, and materials science [66,68–70].

Apart from the practical applications of compact X-ray science, there are growing usages of such compact X-ray sources as platforms to explore fundamental science.

This type of research is particularly prominent with PXR, which was used to study the interaction of free electrons with emerging nanomaterials such as van der Waals (vdW) layered structures [58–60,71,72], of novel compositions such as $WSe_2$, $FePS_3$, and $NiPS_3$ [59]. Another notable experiment demonstrated the generation of PXR using electrons produced by a laser-plasma accelerator, combining this electron source with the PXR mechanism [73]. Most recently, quantum recoil effects in electron radiation, which had been debated and analyzed for many decades [74–78], were first demonstrated experimentally using the PXR platform [79,80], proving the viability of these effects for any electron-radiation process.

> The goal of this review is to examine the underlying physics in the modern coherent sources of hard X-rays. We focus on the mechanism that received substantial recent interest, parametric X-ray radiation (PXR), which is the prominent mechanism of X-ray generation by coherent interactions of free electrons with matter. We review the physics associated with such interactions and the main mechanism by which they produce hard X-rays. The most recent reviews of this mechanism were from 2001 [81] and 2005 [82]. We specifically highlight the advances made since then, as the field made substantial progress in the last decade. Below, we present the state-of-the-art science and applications emerging from this field and build a comprehensive comparison between these mechanisms and the other leading mechanisms for the compact generation of hard X-rays.
>
> *This review provides an entry point to the broader field of X-ray science and its frontier challenges. It serves the wider community interested in compact, tunable, and directional X-ray sources for different applications, and broadens the scope of PXR phenomena, introducing new materials and innovative experimental platforms.*

## 2. Overview of compact X-ray sources

This section reviews the mechanisms of X-ray generations in compact scales. Section 3 focuses on X-ray generation by coherent interaction of free electrons with matter, and section 4 highlights the most recent advances. Based on these recent advances, section 5 draws a roadmap for where the field goes next, toward the realization of a compact source of hard X-rays with sufficient coherence and flux. Section 6 compares the different mechanisms of compact X-ray generation, emphasizing their relative advantages for specific use cases. We conclude the review with an outlook in section 7.

Most mechanisms for producing X-rays rely on energetic free electrons. The term "free electrons" is widely adopted to characterize a beam of electrons after an initial acceleration stage. Throughout the text, "free electrons" serves as an umbrella term that encompasses equivalent expressions found in diverse scientific literature, including "accelerated electrons", "fast electrons", "relativistic electrons", or "swift electrons". Despite the designation "free", these electrons often undergo interactions with various media or with external electromagnetic fields.

### 2.1. Overview of mechanisms for generating X-rays

The different physical mechanisms for producing X-rays can be roughly categorized into four groups (Box 1): (1) Sources based on the incoherent interaction between free electrons and matter, such as bremsstrahlung emitted from an X-ray tube. (2) Sources based on the coherent interaction between free electrons and matter. (3) Sources based on the interaction between free electrons and an external electromagnetic (EM) field. (4) Sources based on the interaction between strong laser fields and matter. This classification, along with descriptions of the different physical mechanisms related to each group, and the relevant metrics for comparison are summarized in Boxes 1-2.

**Box 1 | Physical processes for producing X-rays: classification by interaction types**

X-ray sources differ in their emission spectrum, power, flux, brightness, size, and cost. They can be classified by four interaction types (not accounting for radiation based on nuclear radioactive decay that typically emits gamma rays).

*Sources based on incoherent interaction between free electrons and matter:*

This group includes the X-ray tube, where electrons emitted from a cathode accelerate and impact a target anode, leading to two central emission processes that both rely on local electron interactions with matter that destroy electron coherence: bremsstrahlung and characteristic X-ray radiation. The resulting emission is isotropic and has a broadband spectrum, with a few sharp lines produced by the characteristic radiation. Although operationally simple, with low electron energies and relaxed radiation shielding requirements, the isotropic and broadband emission limit the source brightness and the energy tunability.

*Sources based on free-electron interaction with external electromagnetic (EM) fields*

This group contains synchrotrons, free-electron lasers (FELs) [3,83], and inverse Compton scattering (ICS) [29,30], all relying on periodic electron undulation by external EM fields. In the synchrotron and FEL facilities, the undulation utilizes low-frequency magnetic undulators, while inverse Compton scattering schemes utilize an intense counter-propagating laser beam to undulate the electron [30]. Synchrotron and FEL facilities produce extremely high brightness beams due to the temporal coherence [84], with even higher brightness achieved by electron collective emission in FELs. Nevertheless, these sources are limited by their large size and high cost. Recent proposals for compact X-ray FELs (XFEL) include XFEL Oscillators (XFELOs) [85–87] and free electron lasers driven by optical undulators [88–91].

*Sources based on coherent interaction between free electrons and matter*

This group is based on extended free-electron interactions with matter in a manner that maintains electron coherence throughout its interaction, as in Cherenkov radiation [92,93], transition radiation [94], diffracted transition radiation [95,96], channeling radiation [97–99], coherent bremsstrahlung [100], Smith-Purcell radiation, and parametric X-ray radiation (PXR). While some of these mechanisms are promising for producing quasi-coherent directional X-rays with tunable energy, they are currently limited by heat dissipation, self-absorption of the emitted photons in the matter, and electron scattering. In section 4, we describe recent techniques to mitigate these limitations.

*Sources based on an interaction between intense laser light and matter*

This group relies on external high-intensity laser pulses interacting with matter to produce X-rays [101] and includes high harmonic generation (HHG) [8,10], relativistic flying mirrors [18], and plasma-based X-ray lasers [102]. While these sources can produce coherent beams, most of them are limited to soft X-rays: The HHG spectra are restricted by a cut-off typically reaching up to several hundredths of eVs and rarely to a few keVs [11]. The X-ray plasma laser extension to hard X-rays (>10 keV) is challenging due to the short radiative lifetime (estimated transition times ~1 fs × $\lambda_x^2$, where $\lambda_x$ is the wavelength in angstrom [103]), requiring extremely high pumping intensities. An additional source related to this group is the laser-plasma accelerator (LPA), which is described below.

**Sources based on incoherent interaction between free electrons and matter (the X-ray tube)**

Conventional-anode
Rotating-anode
Liquid-jet anode

*Emission mechanisms*
Bremsstrahlung
Characteristic radiation

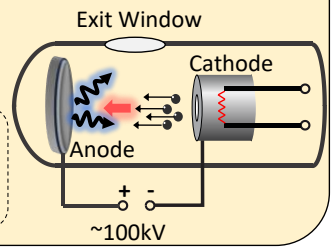

**Sources based on coherent interaction between free electrons and matter**

Cherenkov radiation
Transition radiation
Channeling radiation
Coherent Bremsstrahlung
Smith-Purcell (**)
Parametric X-ray (**)

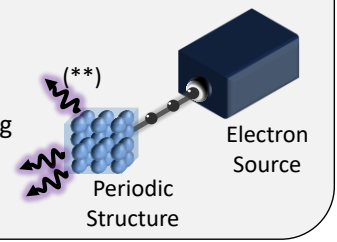

**Sources based on free-electron interaction with EM fields**

Free-electron laser
Synchrotron
Inverse Compton scattering
Betatron

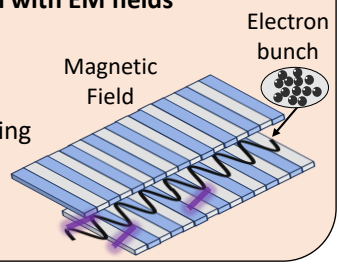

**Sources based on interaction between laser light and matter**

High harmonic generation
Laser plasma accelerators (LPA)
X-ray plasma lasers
Relativistic flying mirrors

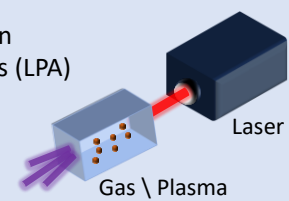

**Box 2 | Physical processes for producing X-rays: scaling laws and spectral range**

The energy scaling of the different X-ray sources unveils a fascinating interplay between emitted X-ray energy and electron energy, with distinctive characteristics shaping their behavior. We focus on five representative physical processes.

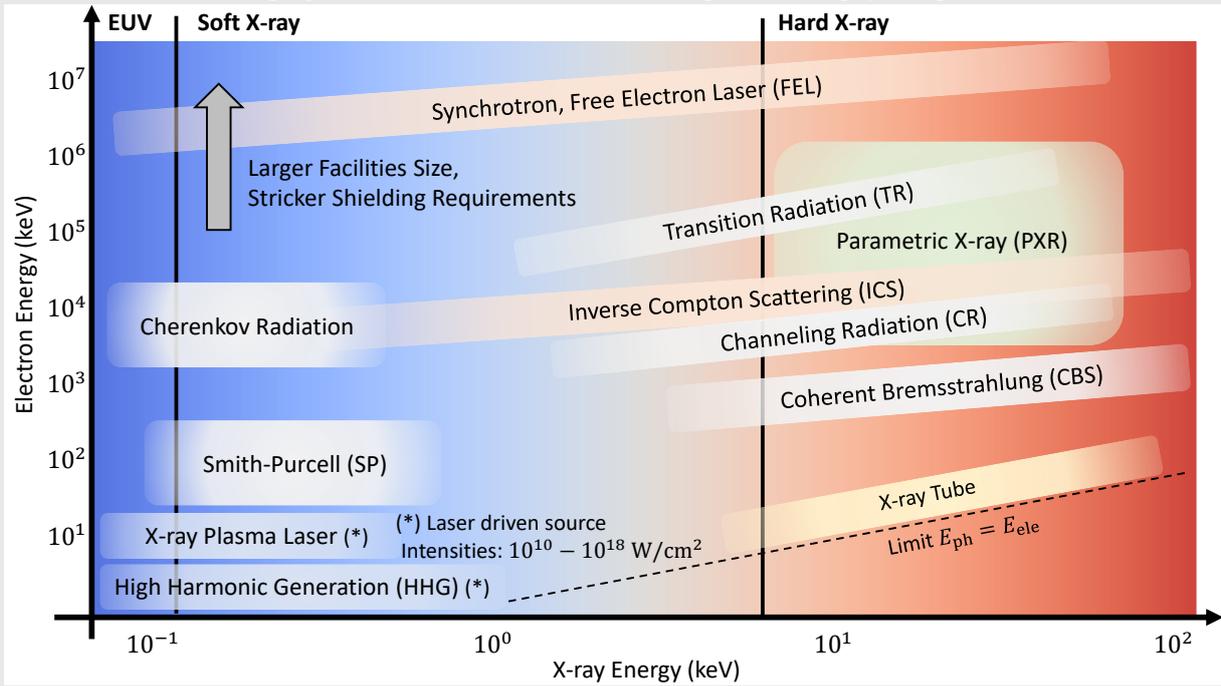

*Undulation mechanisms (synchrotron, free-electron laser, inverse Compton scattering, and coherent bremsstrahlung)*

In these sources, the emitted X-ray wavelength scales quadratically with the electron energy, $\lambda_x \propto \lambda_u \cdot \gamma_e^{-2}$, where $\lambda_u$ is the undulation period and $\gamma_e$ is the Lorentz factor. These sources differ by the undulation period, centimeter-scale ($\lambda_u \sim 1$ cm) for synchrotron and FEL facilities, micrometer-scale arising from the laser wavelength ($\lambda_u \sim 1-10$ μm) for inverse Compton scattering, and angstrom-scale determined by the crystal lattice ($\lambda_u \sim 1$ Å) for coherent bremsstrahlung. Thus, inverse Compton scattering and coherent bremsstrahlung require lower electron energies to achieve the same emitted X-ray energy, compared with synchrotron and FEL facilities. However, this advantage comes at the cost of lower brightness.

*Transition radiation*

Transition radiation is generated when a charged particle passes through an interface between two different media [104]. While transition radiation intensity is maximal in the optical range, its spectrum extends to short wavelengths $\lambda_x \approx \frac{2\pi c}{\omega_p \gamma_e}$, where $\omega_p$ is the plasma oscillation frequency. The linear dependence on the electron Lorentz factor implies that high electron energies are required to produce X-rays.

*Parametric X-ray radiation*

For relativistic electron beams, PXR emission energy is independent of the electron energy but depends on the crystal properties and the emission angle. This allows the production of PXR at the hard-X-rays even with ~10 MeV electron energies, relaxing the requirements on the electron source.

*Channeling radiation*

Channeling radiation is associated with free electrons passing through a crystal while becoming *bounded transversely to the crystal potential* [97]. Confined within the lattice potential well, the emitted photon energy depends on the transition energy between two bound eigenstates of the crystal potential, leading to emission with wavelength dependence of $\lambda_x \propto \gamma_e^{-\alpha}$, where $\alpha$ is typically between 1.5-2, depending on the crystal potential (for example, $\alpha \approx 1.7$ for diamond) [99]. We note that a similar configuration also describes coherent bremsstrahlung and parametric X-ray radiation, however, the physical process is fundamentally different: electrons pass through the crystal *without being bound to its transverse potential* and *interact longitudinally*, resulting in completely different energy scaling laws.

*Soft-X-ray Cherenkov radiation*

Cherenkov radiation is emitted by a charged particle when its velocity in a medium with a refractive index $n$ exceeds the phase velocity of light ($c/n$) [104]. While Cherenkov radiation in the visible and UV spectrum is well known, soft X-ray Cherenkov radiation was historically excluded since the medium refractive index is generally lower than unity in the X-ray spectrum. However, at some inner-shell absorption edges, the refractive index exceeds unity, allowing the generation of Cherenkov radiation in a narrowband region [105]. The Cherenkov radiation in the soft X-ray was demonstrated up to emission energies of ~1 keV [93].

## 2.2. The metrics and scaling laws of X-ray sources

X-ray sources vary in size, beam quality, radiation shielding, safety features, and ease of operation. Among these, beam quality is generally the most significant factor for comparison. The central quality properties of the X-ray beam include transverse coherence (beam emittance) and longitudinal coherence (beam spectral bandwidth). When evaluating X-ray source properties across different methods, it is useful to consider spectral bandwidth, transverse coherence, and longitudinal coherence. These factors are combined into a single metric called brightness (also known as brilliance or spectral brilliance), which allows the comparison of X-ray beam quality from various sources.

The brightness figure of merit for the source is defined as:

$$\text{Brightness} = \frac{\text{photons / second}}{(\text{mrad})^2 (\text{mm}^2 \text{ source area})(0.1\% \text{ BW})}, \quad (1)$$

where BW denotes bandwidth. The brightness expression in Eq. (1) includes four terms. The first denotes the number of photons emitted per second. The second term describes beam collimation, indicating the degree of divergence as the beam propagates, typically measured in milli-radians for both horizontal and vertical directions. The third term addresses the source area's size; a smaller area allows the X-ray beam to be focused to a correspondingly smaller image size, usually measured in $\text{mm}^2$ units. The last term represents the spectral bandwidth. Some X-ray sources produce smooth spectra, while others have peaks at specific photon energies. Therefore, when comparing sources, it is essential to consider the range of photon energies contributing to the measured intensity, often standardized to a fixed relative energy bandwidth (0.1% BW).

### 2.2.1. Comparing X-ray sources by their brightness

**Figure 1** compares the peak brightness of different X-ray sources. The peak brightness metric represents the brightness in a single pulse. Brightness generally depends on photon energy and varies significantly across different X-ray sources. For example, while HHG sources produce high peak brightness in the soft X-ray spectrum [13,23,106], their extension to the hard X-rays is challenging and limited. Third-generation undulators (synchrotrons) have a brightness approximately ten orders of magnitude higher than rotating anodes at the K$\alpha$ line. XFELs achieve even higher peak brightness due to the transverse coherence resulting from coherent emission by micro-bunched electrons [3]. Further improvements are possible by XFELOs, which have the potential to produce longitudinally coherent beams using narrow linewidth mirrors based on X-ray monochromators [2,85–87,107–109]. This significant progress has led to a paradigm shift in experimental X-ray science, allowing experiments that were inconceivable only a few decades ago to be performed routinely.

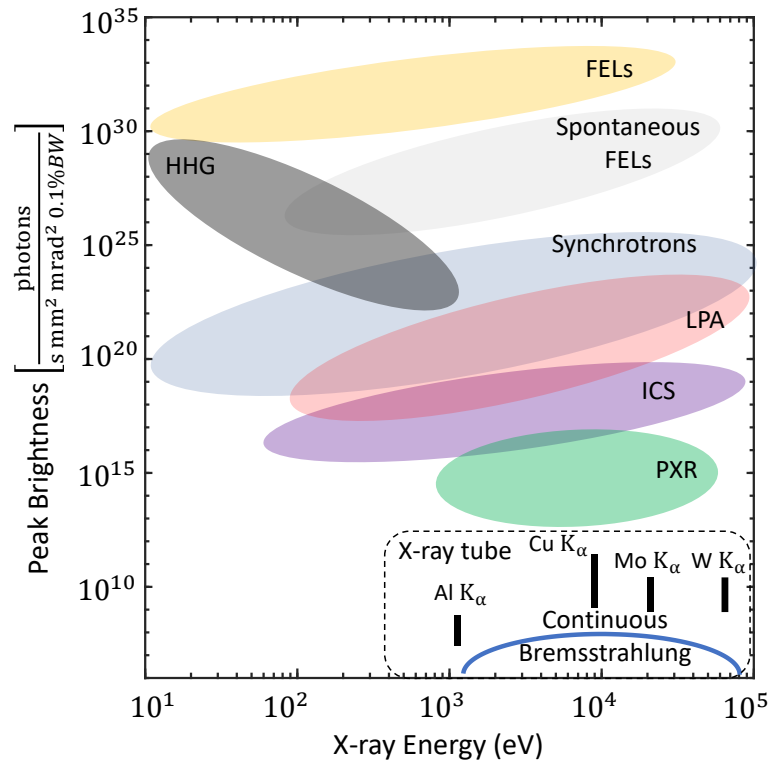

**Figure 1: Peak brightness of different X-ray sources.** The comparison includes the synchrotron and free-electron-laser (FEL) facilities [2,85–87,107], high-harmonic-generation (HHG) [3,13,23,106], laser-plasma-accelerators (LPA) [21,26,110], inverse Compton scattering (ICS) [27–32], parametric X-ray radiation (PXR) [82], and the X-ray tube [65–67]. The peak brightness is defined in Eq. (1). While the brightness metric is the common metric used for comparison between X-ray sources, it is not necessarily the relevant metric for some applications, such as medical imaging that require a relatively large-field-of-view.

Laser plasma accelerators (LPAs) promise to deliver high-brightness X-ray beams in compact setups by accelerating electrons to relativistic energies through the interaction of intense laser pulses with plasma [20,21]. This interaction produces various X-ray radiation mechanisms, including Betatron radiation, Thomson backscattering, and Bremsstrahlung radiation [21,26,110]. Additionally, LPA-accelerated electrons can be injected into a conventional undulator [24]. However, current limitations of LPAs for producing X-rays include limited flux, a broadband energy spectrum, and a limited repetition rate [20,21], which restrict their average brightness.

*2.2.2. Why brightness is not necessarily the relevant metric*

Despite brightness being a common metric, it is not universally suitable for all X-ray applications. Imaging applications, for instance, demand higher X-ray beam flux, with less emphasis on beam emittance due to the necessity of a larger field of view. In contrast, for ultrafast dynamics, high-resolution X-ray spectroscopy, and diffraction applications, the brightness metric is more representative, given the analysis of small-dimension targets. While brightness characterizes source quality for high-resolution applications, flux holds greater significance for imaging applications due to the advantageous larger field of view. Hence, the choice of beam quality metric should align with the target application. A detailed analysis and comparison between PXR, ICS, and the X-ray tube is given in Section 6.

## 2.3. The search for a compact and coherent source of hard X-rays

Despite the widespread use of laboratory-scale X-ray sources, the physical generation mechanisms remained relatively unchanged since the first X-ray tubes, where electrons emitted from a cathode accelerate and impact a target anode in a vacuum tube. The two main mechanisms in X-ray tubes are bremsstrahlung and characteristic X-ray radiation. The typical X-ray tube emission has a broadband spectrum due to the bremsstrahlung radiation, with a few sharp lines produced by the characteristic radiation. This spectrum depends mainly on the anode material and the applied acceleration voltage between the cathode and anode [84]. Recent advances increased X-ray tube brightness by micro-focus sources and liquid-jet anodes [65], enabling new applications in phase-contrast imaging and high-resolution diffraction [69,111]. Notwithstanding these advances, the fundamental limitations in the usage of X-ray tubes remained the same, e.g., its low efficiency, broadband, and isotropic emission.

### 2.3.1. *The challenge of lasing in the X-ray spectrum*

In the past decades, we have witnessed the rise of intense, tunable, and directional X-ray sources in the form of large, expensive synchrotron and free-electron laser facilities [3]. These facilities open the doors to the spectroscopy of material dynamics and biological processes by producing ultrashort X-ray pulses [112]. The coherence of such X-ray sources enables higher-resolution imaging through phase-contrast techniques and next-generation security inspection of microchips [113]. However, the large size and expense of synchrotrons and free-electron lasers have been an obstacle to their widespread adoption in commercial and medical applications.

A long-standing fundamental question at the core of X-ray science is what prevents us from building X-ray lasers based on similar mechanisms as used in conventional lasers in the visible and infrared. Since the development of lasers in the infrared and visible spectral regions in the 1960s [114], there has been a continuous effort to extend the generation of coherent electromagnetic radiation to shorter wavelengths, aiming for the X-ray spectrum. However, the conventional atom-based population inversion approach faces significant challenges when scaling to higher emission energies: (1) *Shorter lifetimes of excited atom-core quantum energy levels:* The radiative lifetime of an X-ray laser transition is estimated to be $\sim 1 \text{ fs} \times \lambda_x^2$, where $\lambda_x$ is the wavelength in angstrom [103]. This extremely short lifetime poses a significant challenge for achieving population inversion. (2) *The energy required for inner core excitation:* The energy required for hard X-ray photon emission is at least four orders of magnitude larger than that for optical photon emission. These two factors result in demanding requirements for the pumping powers necessary to achieve population inversion, a crucial condition for lasing action. Consequently, current X-ray sources based on classical population inversion are not widely accessible, except for some experimental attempts in the 1980s [115], and for ongoing efforts to use X-ray cavities based on crystal Bragg mirrors [116].

*2.3.2. Coherent interactions of free electrons with matter*

What emerged in recent years as an especially promising mechanism for hard-X-ray generation on a compact scale is coherent electron interaction with matter, particularly the hallmark mechanism of parametric X-ray radiation (PXR). The roots of this field date back to radiation effects in the optical spectrum, including the works of Cherenkov (1934) [117], Smith and Purcell (1953) [118], and Fainberg and Khizhnyak [119]. Concurrently, the interaction between high-energy electrons and *crystals* has been investigated since 1934 by von Weizsäcker and Williams [120]. The first coherent emission identified from this type of interaction was coherent bremsstrahlung, analyzed by Heitler [121] and Uberall [122], and channeling radiation, which was predicted theoretically by Kumakhov [123] in 1974, and observed experimentally by Terhune and Pantell [124] in 1975. Recently, it has been demonstrated that a charged particle moving in a channeling regime within a periodically bent crystal can produce undulator radiation, with energies ranging from keV to MeV depending on the crystal's bending period [125].

Within the group of sources based on the coherent interaction between free electrons and matter, parametric X-ray radiation (PXR) is one of the most promising mechanisms for producing a directional, monochromatic, linearly polarized, and tunable hard X-ray source in compact dimensions due to its high spectral yield and large field of view [126]. The desired characteristics of PXR are based on the coherent interaction of free electrons with crystals, arising from phase-matching with the periodic crystal structure [127]. PXR thus differs from the conventional X-ray emission mechanisms of bremsstrahlung and characteristic radiation by having the electron maintain its *coherence* during its interaction and emission. Although PXR has been investigated extensively over the decades, it remained limited in usage due to its low flux. For example, practical mammography imaging requires an X-ray beam flux of $\sim 10^5$-$10^6$ $\frac{\text{photons}}{\text{s mm}^2}$, yet the maximal flux achieved in recent PXR experiments is two orders of magnitude lower than this requirement [128].

In the next section, we review the relative advantages of PXR over other compact X-ray sources. Then, we explore the latest advancements in the PXR field that have led to significant improvements in flux levels (Section 4). These innovations have propelled PXR into the realm of viability for in-vivo imaging applications, opening new possibilities in medical diagnostics and research. Finally, we outline a roadmap detailing the steps necessary to achieve a fully realized PXR source (Section 5). This plan would serve as a guide for researchers and engineers, paving the way for implementing this cutting-edge technology for commercial use.

## 3. Background on parametric X-ray radiation

One of the most promising mechanisms for producing quasi-coherent and tunable hard X-rays in compact scales is PXR, which was experimentally demonstrated for the first time in 1985 [129]. The last comprehensive reviews of PXR were conducted two decades ago [81] [82], but the field has seen a significant revival since then, with fundamental experimental discoveries, refined theoretical models, and new applications [79,80,130–137]. Significant engineering progress was made toward realizing a compact implementation of the PXR mechanism [137,138]. Of particular importance are X-ray phase-contrast imaging applications, which have been demonstrated using PXR [128,139–141]. These experiments were shown in large facilities but proved the mechanism that can be implemented on compact scales.

The next phase in the development of a viable, compact, and widespread PXR source for imaging applications has recently become possible due to three central factors: (1) Progress in synchrotron and FEL facilities promoted the miniaturization of relativistic electron acceleration structures that support high brightness, high repetition rate, and high-average current [29,142,143]. (2) Demonstration of PXR imaging applications, such as K-Edge imaging, phase-contrast imaging using differential-enhanced imaging (DEI), and computed tomography (CT) [128,139–141]. (3) Theoretical contributions discovered PXR geometries that resolved long-standing limitations and increased the spectral yield [144–147].

This section reviews the central aspects of PXR theory, focusing on its superior yield and beam quality in the X-ray spectrum compared to other electron-matter interaction processes. We begin with the fundamental properties and present the kinematical and dynamical theory of PXR, followed by a discussion of its emission properties, including spatial dispersion, polarization, yield, diffraction efficiency, and spectral linewidth. We also address unique aspects of free-electron interaction with matter, such as electron scattering effects, thermal load on the target crystal, and self-absorption of emitted PXR photons. While PXR is the primary focus of this review, the insights and advancements discussed apply to the other X-ray sources based on coherent electron-matter interactions.

### 3.1. Basic mechanism and motivation for parametric X-ray radiation

PXR is produced from the interaction between relativistic electrons and a periodic crystalline structure (**Figure 2**) [82]. It has several desired properties, which can serve various applications. (1) The X-ray spectrum has a narrow linewidth, i.e., it is quasi-monochromatic. (2) The X-ray photon energy can be tuned by crystal orientation, composition, and strain. (3) The X-ray photon energy is practically independent of the incident electron energy in the relativistic regime. (4) The X-ray beam has low spatial divergence, which can be shown to be inversely proportional to the incident electron energy ($\gamma_e^{-1}$) for a wide range of parameters (see section 3.4.1) [148].

Compared to other X-ray sources based on interactions with free electrons, such as transition radiation and synchrotron radiation, hard-X-ray generation from PXR requires a much lower electron-beam energy. For example, to produce 10 keV X-ray photons, synchrotron sources require electron beam energy of a few GeV, while transition radiation requires a few hundredths MeV (Box 2). PXR, on the other hand, occurs even at energies below 10 MeV [132,149]. This advantage is directly attributed to the nanoscale and sub-nanoscale periodicities encountered by the electron in the PXR scheme, as opposed to the centimeter scale periodicities typically found in traditional undulators. The low electron energy makes the PXR source considerably more compact and less expensive than synchrotron sources. At the higher end of electron energies, the

PXR mechanism is still applicable, and indeed, tunable PXR was observed with photon energies up to 400 keV from electron beams of 1.2 GeV [150,151].

The PXR source spectral yield (i.e., the average number of photons produced per electron) is up to four orders of magnitude greater than other X-ray sources, such as bremsstrahlung, transition radiation, and coherent bremsstrahlung [126]. The PXR emission spectral linewidth is narrow and proportional to $\propto \gamma_e^{-1}$ at moderate and high electron energies, which is suitable for phase-imaging applications [152]. In contrast, coherent bremsstrahlung and channeling radiation have much higher spectral linewidth [34]. These sources' linewidth is inversely proportional to the number of undulation periods. Since the electron energies required for these sources in the X-ray spectrum are below a few tens MeV (Box 2), the electron scattering in the crystal is significant, limiting the number of effective undulation periods to the order of ~10 (i.e., their linewidth is ~10%) [34].

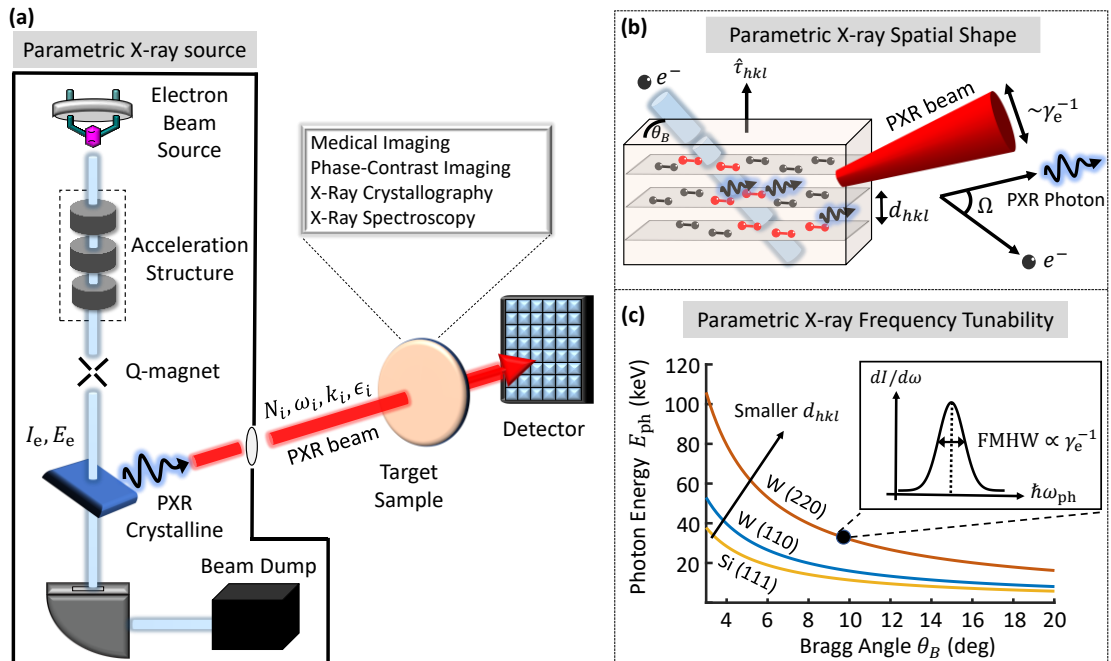

**Figure 2: Parametric X-ray (PXR) source: spatial shape, and energy tunability**. **(a)+(b)** The PXR source scheme. A collimated electron source beam impacts a crystal and induces polarization currents on the target material atoms. Each excited atom can be treated as a radiating dipole. When the Bragg condition of constructive interference between the dipoles array holds, an intense, directional, and quasi-monochromatic X-ray beam is emitted at a faraway angle from the electron velocity direction. **(b)** The PXR spatial emission. The incident electron beam impacts the crystal Bragg plane $\hat{\tau}$ with an angle $\theta_B$. The PXR photon is produced

with an angle $\Omega$ relative to the electron trajectory. The Bragg condition holds for $\Omega = 2\theta_B$. The PXR photons are emitted within an angular divergence of $\theta_{\text{ph}}^2 = \gamma_e^{-2} + (\omega_p/\omega)^2$, where $\omega_p$ is the plasma frequency and $\omega$ is the emission PXR frequency. For most applications of PXR, $\gamma_e \ll \omega/\omega_p$, such that the beam divergence can be approximated by $\theta_{\text{ph}} \sim \gamma_e^{-1}$. (c) PXR frequency tunability. The PXR photon's energy is tuned by altering the Bragg angle and choosing the Bragg plane. When reducing the interplane distance $d_{\text{hkl}}$, the emitted photon energy increases for a fixed Bragg angle. The typical spectral linewidth of the PXR can be as low as ~1%.

Moreover, channeling radiation, coherent bremsstrahlung, and transition radiation emit in the forward direction, parallel to the electron velocity direction. If the target material is thick, the emitted photons are self-absorbed in the material, limiting the source yield. On the other hand, PXR emits at a large angle relative to the electron velocity, enabling special geometries where self-absorption effects are less considerable. In addition, the large emission angle eliminates the need for a strong magnetic field to separate the electrons from the X-rays and minimizes the bremsstrahlung background radiation.

Except for the energy tunability by crystal rotation, the PXR radiation has other characteristics that make it a promising physical mechanism for a compact X-ray source. The PXR emission is directional, polarized, and partially coherent, as discussed in the next section. Furthermore, its polarization and spatial shape can be designed and shaped (**Figure 5**(a)) [153]. For instance, the PXR beam can have either a radial polarization with a circular shape peak or a linear polarization with two lobes shape, depending on the emission angle. The PXR spectrum is independent of the incident electron energy for relativistic electrons, enabling the integration with high energy spread electron sources [73]. PXR radiation angle can be as large as 180 degrees (backscattering), and it has no theoretical limits for the incident electrons' energy.

## 3.2. Milestones in the development of parametric X-ray radiation

**Figure 3** summarizes the main milestones in the development of PXR since the theory's establishment at the beginning of the 1970s [154–159]. After the first observation of PXR in 1985 [129,160], its basic mechanism was tested and analyzed experimentally, leading to additional refining of the theory in the 1990s and 2000s [81,131,134,161–173]. Since the beginning of the 2000s, the focus has moved to PXR applications, especially for X-ray imaging [141,174–179] (Table 1 and Box 3), as well as for pulsed PXR sources [180], electron beam diagnostics [95,181,182], PXR lens focusing by bending crystals [71,72,183], calibration of X-ray space telescopes [184], detection of nuclear materials [185], and measurement of the crystalline grains size in polycrystals [186].

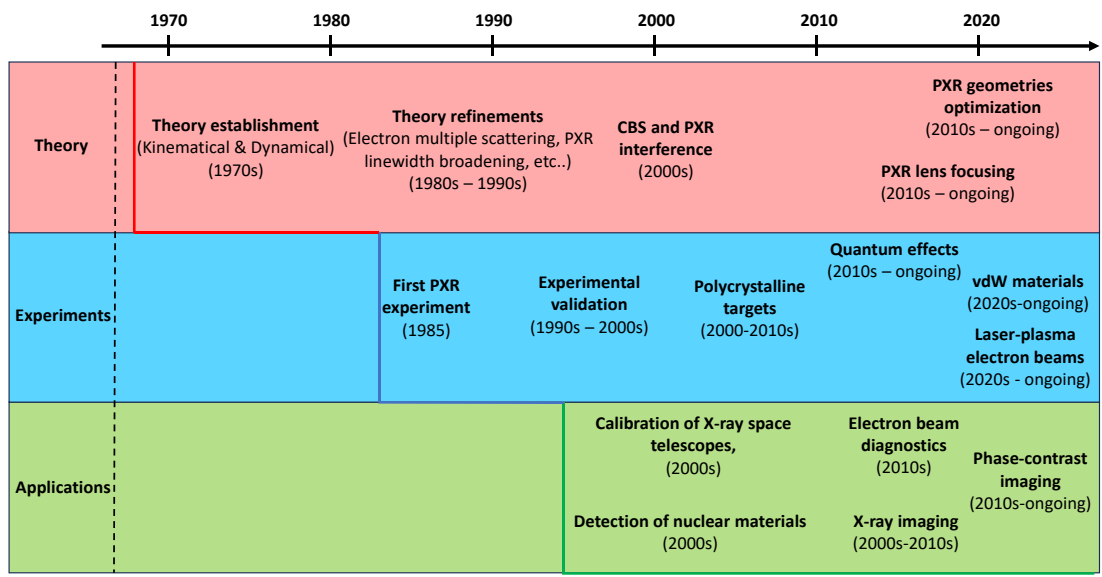

**Figure 3: Timeline of developments in the field of parametric X-ray radiation (PXR)**, from its inception to the most recent ongoing efforts.

The first experimental realization of PXR was done in 1985 by V. G. Baryshevsky et al. [129,160] using a 900 MeV electron beam from the Tomsk synchrotron to produce a 6.96 keV PXR from a diamond crystal. Since then, numerous studies have been conducted to characterize PXR from different materials: silicon (Si) [162], germanium

(Ge) [187], molybdenum textured polycrystal (Mo) [188,189], highly oriented pyrolytic graphite (HOPG) [190–192], diamond [193], tungsten (W) [194], copper (Cu) [195], aluminum (Al) [196], lithium fluoride (LiF) [174,175], and gallium arsenide (GaAs) [197]. A detailed review of experiments conducted before 2005 can be found in [198]. Later years have also characterized PXR from novel materials such as various polycrystalline solids [133,199,200], multilayer X-ray mirrors [201], van-der-Walls (vdW) materials [59,60], and even powders [132], instead of the traditional monocrystal bulk solids. An additional focus in the last years has been on optimizing the PXR geometry [144–147] and demonstrating quantum effects [79,80].

The first PXR experiments were performed in synchrotron facilities with electron beam energies of hundreds of MeV [129,160,162]. Later experiments used linear accelerators with electron energies of tens of MeV [152,175,176,178,179,202]. PXR was recently also demonstrated with the typically lower quality electron beams produced by laser plasma, enabling prospects with plasma-based electron sources [73]. Certain recent proof-of-concept studies were demonstrated using electron microscopes of tens to hundreds of keV [58–60,80].

These recent theoretical and experimental contributions pave the way toward a compact PXR source, using moderate electron energies [59,203,204]. At the lower energies, interference between PXR and coherent bremsstrahlung becomes important. However, the yield and brightness in these lower energy regimes are significantly lower compared to PXR with relativistic electrons [203,204]. Consequently, sources aiming at X-ray applications (rather than fundamental demonstrations of novel concepts) focus on regimes of relativistic electron, where PXR dominates over coherent bremsstrahlung. In the next section, we review the central aspects of PXR theory, the experimental progress, and recent application achievements of X-ray imaging.

**Box 3 | Applications of parametric X-ray radiation**

*Applications of high-coherence X-ray sources*

PXR is a prospective mechanism for producing quasi-coherent X-ray radiation. High-coherence X-ray sources are prospective for numerous applications, from medical imaging to high-spatial-resolution imaging of biological samples and nanocrystals. The most notable applications are listed below.

**Phase-contrast imaging** utilizes the phase shift that occurs during X-ray transmission and scattering by an object [68]. Extracting the X-ray beam's phase shift enables the creation of high-contrast images, particularly beneficial for visualizing details in specimens with weak absorption contrast, such as soft tissues like lungs and breast tissue (Figure (a)).

**K-edge imaging** enhances element contrast by using the significant differences in the sample's photo-electric attenuation coefficient above the K absorption edge [205]. Monochromatic X-ray beams slightly below and above the K-edge produce two images with distinct intensity maps, facilitating the detection of fine structures and improving overall image contrast (Figure (d)).

**Coherent diffraction imaging (CDI)** is a powerful technique for reconstructing high-resolution structures of samples [206]. CDI enables the extraction of both amplitude and phase information from non-crystalline samples, expanding the range of studied specimens to those impossible to crystallize, such as various biological samples [207], providing valuable insights into the nanoscale and atomic structures of diverse materials.

Besides imaging applications, coherent X-ray sources can be beneficial for X-ray scattering and spectroscopy applications, such as small angle X-ray scattering (SAXS) [208], X-ray absorption spectroscopy (XAS), and X-ray absorption fine structure (XAFS) [209].

*Demonstration of imaging using parametric X-ray radiation*

While the listed applications above operate optimally in the large and expensive synchrotron and FEL facilities, a substantial effort is made to produce a compact, high-coherence X-ray source. In the last two decades, PXR sources have been demonstrated for imaging applications. Two labs have shown the PXR feasibility as a compact and tunable source for imaging – the first group is from Rensselaer Polytechnic Institute (RPI) (2002-2009) [174,175,195], and the second group is from LEBRA, Nihon University which is active since 2004 [141,177–179].

In these studies, images capturing the absorption of computer chips and animals were obtained (Figure (a)). Furthermore, the experiments demonstrated phase-contrast imaging (Figure (c)) and 3D tomography (Figure (b)), successfully capturing K-Edge subtraction CT images (Figure (d)) [140]. These results suggested that PXR has spatial coherence and is a suitable X-ray source for imaging.

Despite the significant progress made in these experiments, they were still limited by a requirement for a long exposure time (~tens of seconds) due to insufficient flux levels (Table 1).

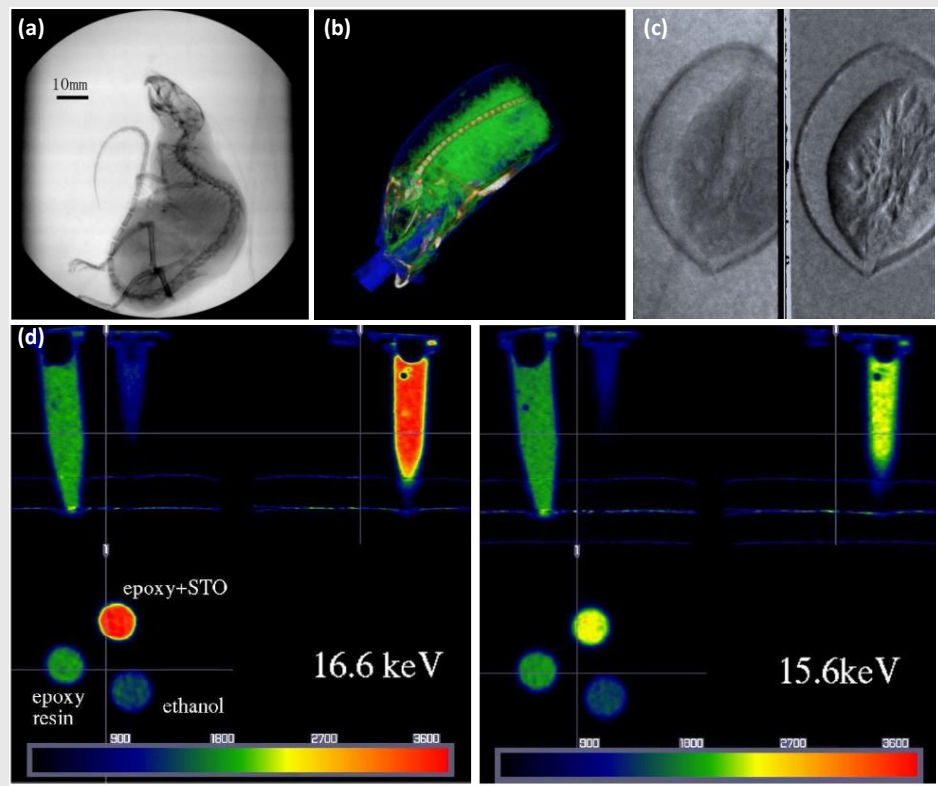

**Snapshots from PXR imaging experiments.** (a) An absorbing X-ray image of a mouse was observed using a 25.5 keV PXR beam. (b) 3D tomography for a raw fish sample. The tomogram was reconstructed from 180 projection images using a 17.5 keV PXR beam. (c) Absorption-contrast (left) and phase-gradient (right) images from 34 keV PXR beam. (d) K-Edge subtraction CT image taken with 16.6 and 15.6 keV PXR beams. (a) – (c) taken from [128] and (d) taken from [140].

**Table 1:** PXR experiments for imaging applications. The electron source, target crystal, X-ray emission spectrum, and target sample dimensions are compared.

| | Parameter | Y. Hayakawa et al. [179] [152] | B. Sones et al. [174–176] |
|---|---|---|---|
| **General** | Year | 2004-current | 2002-2009 |
| | Facility | LEBRA, Nihon University | Rensselaer Polytechnic Institute |
| **Electron Source** | Energy | 100 MeV | 56 MeV |
| | Energy spread | $\leq 1\%$ | $\leq 15\%$ |
| | Electron pulse duration | 4-5 µs | 30 ns |
| | Peak current | 120-135 mA | 1.5 A |
| | Repetition rate | 2-5 Hz | 400 Hz |
| | Average beam current | 1-5 µA | 0.01 - 6 µA |
| | Normalized emittance | ~15π mm mrad | Not reported |
| | Electron beam size on target (diameter) | 0.5 – 2 mm | ~1 cm |
| **Target Crystal** | Materials | Silicon | Lithium fluoride (LiF) |
| | Thickness | 200 µm | 500 µm |
| | Geometry | Bragg / Laue | Bragg / Laue |
| | Bragg Angle | $5.5° - 30°$ | $15°$ |
| **X-ray Photon** | Photon Energy | Si (111) - 4-20 keV<br>Si (220) - 6.5-34 keV | 6 – 35 keV |
| | Total X-ray photon rate (photons/s) | $\sim 10^7$ | $\sim 10^6$ |
| **Target Sample** | Distance from PXR source | ~ 10 m | ~ 3 m |
| | Beam diameter on target | ~100 mm | ~3 mm |
| | Total X-ray photon flux (photons/mm$^2$/s) | $\sim 10^3$ | ~ 75 |

### 3.3. Fundamentals of parametric X-ray radiation

PXR radiation occurs when a relativistic charged particle passes through an aligned crystal (**Figure 2**). In this review, we discuss an electron source beam, but other charged particles, such as protons, exhibit similar phenomena [135,210–212]. The PXR production mechanism has been studied since 1970 by Ter-Mikaelian [159,213], Baryshevsky and Feranchuk [155,158], and Garibyan and Yang [154,156]. The most immediate feature that made PXR stand-out relative to other X-ray emission mechanisms was a sharp X-ray emission at a large angle relative to the electron motion direction. This large emission angle contrasts with other X-ray radiation sources, such as bremsstrahlung and transition radiation, that emit nearly parallel to the electron motion direction.

The PXR emission is also spatially narrow and confined to a cone shape inversely proportional to the Lorentz factor of the electron $\gamma_e^{-1}$ at moderate electron energies (**Figure 2**(b)). Baryshevsky and Feranchuk gave this radiation the name PXR by analogy to the optical radiation, considered by Fainberg and Khizhnyak [119] but additional names are also in use: dynamical radiation, resonance radiation, quasi-Cerenkov radiation, or dynamical Cerenkov radiation [164].

*3.3.1. The X-ray generation mechanism*

Several equivalent descriptions exist for the PXR phenomenon. In one description, a collimated electron source beam impacts a crystal and induces polarization currents on the target material atoms. Each excited material atom acts as a radiating dipole. When the Bragg condition of constructive interference between the dipoles array holds, an intense, directional, and quasi-monochromatic X-ray beam is emitted at a large angle relative to the electron velocity direction (**Figure 2**). The maximum PXR production is when two conditions are satisfied simultaneously: the Smith-Purcell condition for the dipoles parallel to the electron trajectory axis and the requirement of transverse plane dipoles' constructive interference [153,171].

An equivalent description of the PXR phenomenon is the diffraction of the electron's virtual photon field by an array of atoms in the crystal. The diffracted virtual photons appear as real photons at the Bragg angle corresponding to the diffraction of X-rays, i.e., the virtual photons diffract from the crystal planes in the same way as real photons. The Bragg law governs the X-ray diffraction conditions and relates the photon energy, the interplane d-spacing between crystal planes, and the incident angle between the photons and the diffraction plane. Consequently, continuously tunable PXR production is possible with the rotation of the target crystal (**Figure 2**(c)).

Two configurations are commonly employed for generating PXR, analogous to those in X-ray crystallography: Bragg and Laue geometries (**Figure 4**). The two configurations differ by the direction of X-ray emission relative to the crystal "front" surface through which the electron impinges the crystal. In Bragg geometry, the PXR reflection is emitted from the front surface of the crystal [141,152,177–179,214,215] (top in **Figure 4**), whereas in Laue geometry, it is emitted from the rear surface [173,216] (bottom in **Figure 4**). The change in emission angle arises from the specific choice of families of crystallographic planes with which the electron interacts.

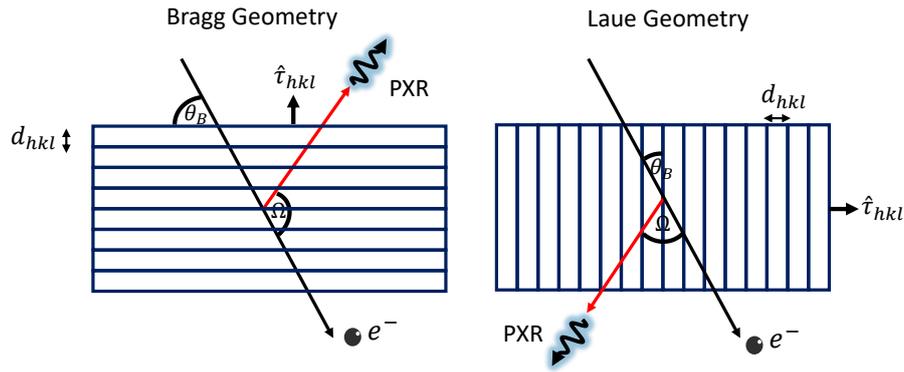

**Figure 4: The two configurations of parametric X-ray radiation (PXR): Bragg and Laue geometries.** The electron beam impinges the crystal through the front surface and exits through the rear surface. In Bragg geometry, the PXR reflection is emitted from the front (top) surface of the crystal, while in Laue geometry, it is emitted from the rear (bottom) surface.

*3.3.2. The dynamical and kinematical theories*

The theoretical framework of PXR can be divided into the kinematical theory and the dynamical theory (similar to the division in the theory of X-ray diffraction [84,217]). The PXR dynamical theory was developed by Baryshevsky and Feranchuk [155,158], Garibyan and Yang [154,156], and Caticha [164,218] and considers all the PXR multiple scattering effects, including refraction, extinction, and interference effects, which alter the shape and width of the PXR peaks. In contrast, the simplified kinematical theory ignores these effects, as done in the description of Ter-Mikaelian [159,213], Feranchuk and Ivashin [161], and Nitta [163,165,166,169], and was recently rederived for heterostructures [153]. The kinematic theory is based on the framework of classical electrodynamics, while Nitta's work provided a quantum derivation that aligns with the classical predictions.

The dynamical theory of PXR extends the kinematic theory but is more challenging to apply in practice. Generally, the dynamical theory provides the most accurate predictions for the total radiation intensity. The differences between the dynamical and kinematic theories are most significant near the Bragg peaks, particularly in thick crystals, where refraction, extinction, and interference effects should be considered [219]. While the kinematical theory is valid for thin materials below the extinction length ($L_{\text{ext}} \sim 1$ µm) [82], ongoing efforts aim to define the boundaries where the kinematic theory remains accurate [134,220]. Research into these boundaries is particularly interesting in specialized PXR geometries, such as bent crystals, which create overlaps between Bragg and PXR peaks, thus altering the kinematic theory's validity and potentially revealing new resonances detectable only by the dynamical theory [221].

While more precise PXR experiments can help clarify these boundaries, extensive research during the 1990s and 2000s refined the PXR kinematic theory to better align with experimental results for thicker materials [175]. Consequently, the refined PXR kinematic theory has become the most commonly used in practice. Throughout this review, we will use this refined PXR kinematic theory.

*3.3.3. Ultra-relativistic electron beams*

The PXR emission energy is closely related to the Bragg law governing the diffraction of an incident X-ray beam from a crystal. The Bragg law relates the incident X-ray beam energy $\hbar \omega_B$ and the Bragg angle $\theta_B$ between the incident X-ray momentum vector and the reflective crystallographic plane [84]:

$$E_B = \hbar \omega_B = \frac{\pi \hbar c}{d_{\text{hkl}}} \frac{1}{\sin(\theta_B)}, \qquad (2)$$

where $d_{\text{hkl}}$ is the d-spacing of the Bragg plane corresponding to Miller indices (hkl).

Similarly, the expression for the PXR emission energy as a function of the Bragg polar angle $\theta_B$, for ultra-relativistic electrons, can be derived from the energy and momentum conversation lows in a crystal [159]:

$$E_{\text{PXR}} = \hbar \omega_B = \frac{2\pi \hbar c}{d_{\text{hkl}}} \frac{\sin \theta_B}{1 - \sqrt{\epsilon}\beta \cos \Omega}, \qquad (3)$$

where $\beta = v/c$ is the normalized velocity of the electron ($\beta \approx 1$ for ultra-relativistic electrons) and $\epsilon$ is the constant part of the medium permittivity ($\epsilon \approx 1$ for hard X-rays). $\Omega$ is the emission polar angle of the PXR photons relative to the electron beam, with Bragg's law imposing practical phase matching satisfied at the polar angle $\Omega = 2\theta_B$, around which the maximum PXR intensity is obtained. Eq. (3) can be derived from Huygens' construction [173] and can be interpreted as the expression for a Doppler frequency in a medium [213].

Both Bragg diffraction (Eq. (2)) and PXR emission frequency (Eq. (3)) are obtained via phase-matching arguments. The Bragg diffraction considers an incident X-ray beam (i.e., an incident photon), while the PXR diffraction considers an electron moving at a constant velocity as the source of the electromagnetic field. This assumption of constant velocity is a common classical assumption that holds for a broad range of parameters, enabling us to reach analytical results. Thus, the Bragg law (Eq. (2)) and the PXR energy (Eq. (3)) are closely related: the Bragg frequency in Eq. (2) is obtained for an incident X-ray beam corresponding to $\beta\sqrt{\epsilon} = 1$ and an observation angle of $\Omega = 2\theta_B$, rather than $\beta\sqrt{\epsilon} < 1$ for the incident electrons. In other words, the constructive interference conditions for Bragg law and the PXR emission are similar, with the main difference being the slightly lower velocity of an incident electron compared to an X-ray photon. As a result, the Bragg frequency is slightly higher than the PXR frequency [134,222]. Further details and comparison between the Bragg frequency and PXR frequency can be found in [162,173,216,222,223].

An additional constraint limits the azimuthal angle of emission, arising from transverse (relative to the electron motion direction) phase matching with the crystal lattice. For instance, in a hexagonal lattice, there is an azimuthal symmetry of $\pi/3$ [153]. The emission is confined around discrete emission angles with an opening angle proportional to $\propto \gamma_e^{-1}$ (at moderate and high incident electron energies), as shown in **Figure 2**(b). This relation allows PXR energy tunability in experiments by rotating the PXR crystal, i.e., altering the $\Omega$ and $\theta_B$ angles [175]. In particular, the PXR photons' energy is almost independent of the incident electron energy for relativistic electrons with energy above ~5 MeV. At these highly relativistic electron energies, the photon energy is determined solely by the spacing between the crystal planes and the experimental geometry that determines the angles.

*3.3.4. Moderately relativistic electron beams*

The first decades of PXR studies focused mostly on ultra-relativistic electron energies (a few tens of MeV and above) in synchrotrons, storage ring facilities, and linear accelerators. Recent years have shown a significant growing interest in moderate electron energies of only a few hundredths keV and even below, with PXR being observed down to a few tens of keV [58,60,71,80]. Such moderate electron beam energies would open many opportunities, including compact electron sources with reduced source shielding requirements.

The expression for the PXR emission energy as a function of polar angle, for moderately relativistic electrons, is similar to Eq. (3) [159]. Unlike the similarity of the polar-angle dependence, other properties of PXR differ substantially from the ultra-relativistic regime. These properties include the emission spectrum, the spatial shape of the radiating beam, and its angular distribution. Specifically, the emission is not confined to discrete directions but instead spreads across a wide range of angles. This emission spreading occurs because there are no phase-matching conditions imposed along the transverse plane (perpendicular to the electron motion direction). Another difference from the ultra-relativistic regime is that in the moderately relativistic regime the interference between PXR and coherent bremsstrahlung (CBS) becomes considerable [203,204], as was studied and observed experimentally in [224,225].

The same polar-angle dependence of Eq. (3) applies to a wider family of electron radiation phenomena besides PXR, including coherent Bremsstrahlung and Smith-Purcell radiation [153]. The latter is emitted from electrons passing by a periodic optical structure and satisfying phase matching along their direction of motion. Smith-Purcell radiation was observed in the radiofrequency [226], optical [227], terahertz [228], and more recently ultraviolet [229] spectra; its analogy to PXR is characterized in [153]. This connection highlights the universality of PXR physics.

Interestingly, quantum-recoil effects can cause deviations from Eq. (3) [230] in any of the mechanisms it applies to. This universal quantum effect was recently observed for the first time using a PXR experiment [80], using electron energies of tens of keV. In such a regime, the output X-ray energy can deviate substantially from Eq. (3), allowing greater versatility in controlling the X-ray spectrum [80]. In this review, however, we focus on the regimes most prevalent in experiments, where Eq. (3) accurately predicts PXR.

### 3.4. Emission characteristics of parametric X-ray radiation

In this section, we present the PXR spatial shape and dispersion, polarization, and yield for different crystal materials. The PXR yield depends on several factors, including the target material, the crystal geometry, the diffraction efficiency, and the thermal load on the crystal. In the framework of the kinematical theory, the photon distribution emitted from a single electron is given by [172]

$$\frac{dN_{\text{PXR}}}{d\theta_x d\theta_y} = \frac{\alpha}{4\pi} \frac{\omega_B}{c \sin^2 \theta_B} f_{\text{geo}} \chi_g^2 e^{-2W} N(\theta_x, \theta_y), \qquad (4)$$

where $\alpha$ is the fine-structure constant, $\omega_B$ is the emitted PXR photon energy, $c$ is the speed of light, $\theta_B$ is the Bragg angle, $e^{-2W}$ is the Debye-Waller factor which captures thermal effects, $\chi_g$ is the Fourier expansion of the electric susceptibility as a function of the reciprocal vector $\boldsymbol{g}$, describing the diffraction efficiency (Eq. (7)), $N(\theta_x, \theta_y)$ is the PXR angular dependence (Eq. (5)), and $f_{\text{geo}}$ is the geometrical factor that describes the PXR photon self-absorption during the emission process (Eq. (10)).

The PXR photon energy ($\omega_B$) and the Bragg angle are related by the condition for constructive interference between the material's dipoles (Eq. (3)). The PXR angular dependence $N(\theta_x, \theta_y)$ is given by

$$N(\theta_x, \theta_y) = \frac{\theta_x^2 \cos^2(2\theta_B) + \theta_y^2}{\left(\theta_x^2 + \theta_y^2 + \theta_{\text{ph}}^2\right)^2}, \tag{5}$$

where $\theta_x$ is the angle in the diffraction plane, $\theta_y$ is the angle perpendicular to $\theta_x$ in the diffraction plane and $\theta_{\text{ph}}^2 = \gamma_e^{-2} + (\omega_p/\omega)^2$, where $\omega_p$ is the plasma frequency of the material. Eqs. (4)-(5) describe the radiation yield for PXR emission near the resonant Bragg angles in both the forward and backward hemispheres; these near-Bragg conditions are the most favorable for efficient PXR. Studies on the PXR yield at large deviations from the Bragg angles are discussed in [223]. The specific shape of the PXR emission perpendicular to the incident electron velocity vector is detailed in [231]. Eq. (5) describes the spatial profile in the diffraction plane, as shown in **Figure 5**(a) for different Bragg angles.

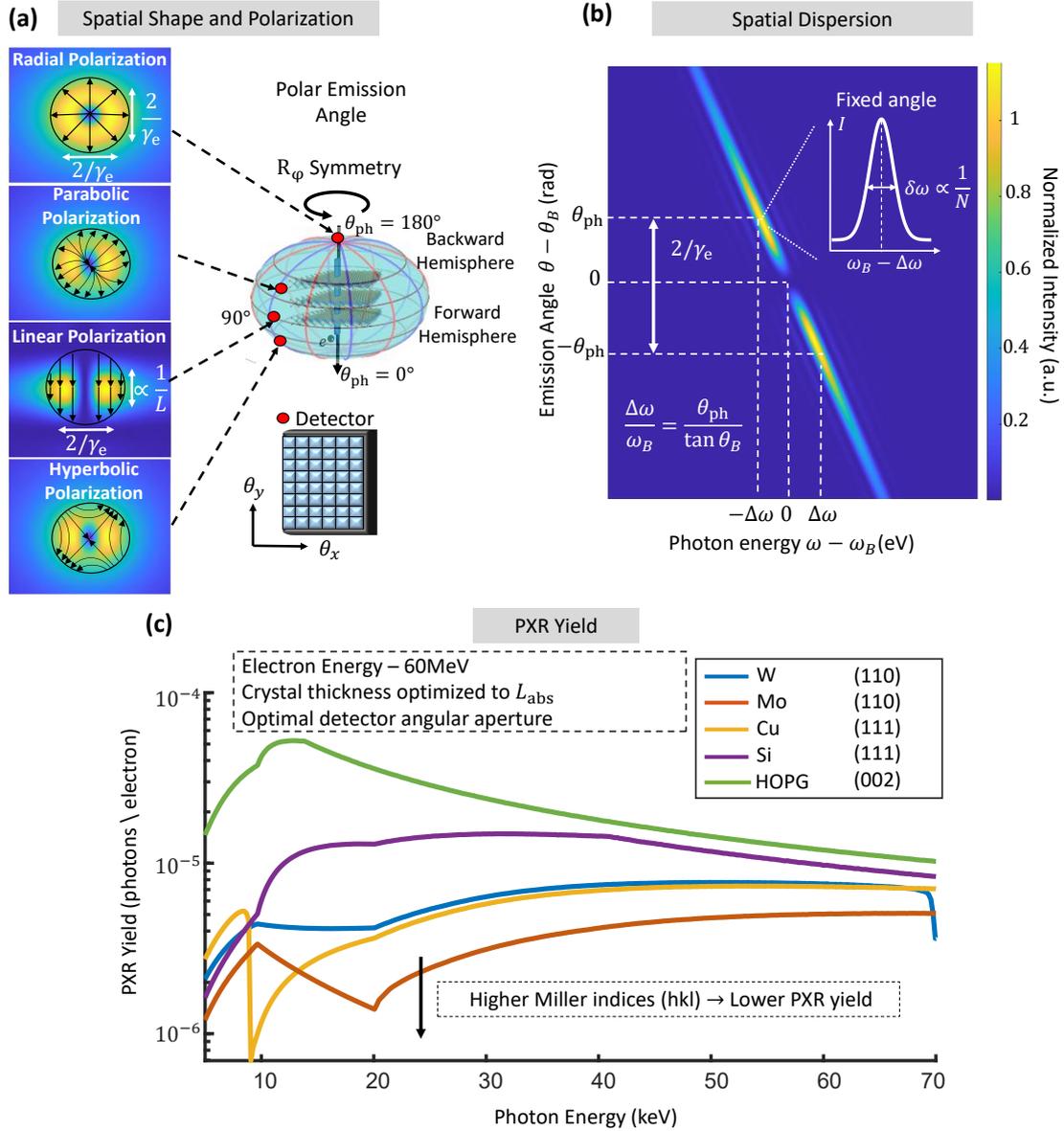

**Figure 5: Parametric X-ray angular distribution, polarization, spatial dispersion, and yield**. **(a)** PXR spatial shape (i.e., angular distribution) and polarization. The PXR spatial shape, as described by Eq. (5) can be either a donut shape or a two-lobes shape, depending on the emission angle, whereas the polarization can be radial, linear, parabolic, or hyperbolic. The yield and polarization are shown for the forward and backward hemispheres, as well as for the perpendicular emission. **(b)** PXR spatial dispersion shape. The angular distribution emission width (HWHM) scales as $\sim \gamma_e^{-1}$ at moderate and high impinging electron energies. This type of angular emission is often called PXR reflection [173]. The polar-angle dependence of the emitted energy is given by Eq.(3). When looking at a fixed emission angle, and at an ideal experimental resolution $\Delta\theta_d = 0$, the intrinsic spectral linewidth is proportional to $\delta\omega/\omega \propto 1/N$, where $N$ is the number of crystallographic planes. **(c)** PXR yield for tungsten (W), Molybdenum (Mo), Copper (Cu), Silicon (Si), and highly oriented pyrolytic graphite (HOPG), denoting the chosen d-spacing $d_{hkl}$ of each material. The calculation assumes an ultra-relativistic electron beam (60 MeV). The crystal thickness is optimized to the absorption length of each of the materials for each PXR energy. For smaller d-spacing, the yield decreases due to lower momentum transfer efficiency (Eq. (8)). The parameters of each material are presented in Table 2. The PXR yield is up to four orders of magnitude greater than other X-ray sources, such as bremsstrahlung, transition radiation, and coherent bremsstrahlung [126].

*3.4.1. The density effect in PXR*

The PXR angular dependence, described in Eq. (5), shows a saturation for electron energy exceeding $\gamma_e \geq \omega/\omega_p$. This saturation phenomenon, attributed to the density effect, arises due to corrections of PXR from ultra-relativistic particles [216,232]. Two central explanations exist for this phenomenon; both predict the same behavior. The first explanation is analogous to the Fermi density effect of ionization energy losses of a fast particle in a condensed medium. In dense media, numerous atoms lie between the incident electron and a far atom in the perpendicular plane to the electron trajectory. These atoms, influenced by the fast particle's fields, produce perturbing fields at the chosen atom's position, modifying its response to the fields of the fast electron. Essentially, each atom is affected by its neighbors, altering its polarizability relative to its free-space value [153]. A different approach for describing this phenomenon was proposed in several works and attributed to the Ter-Mikaelian longitudinal density effect [148,159,216,232]. This approach explains the phenomenon using the formation length [213,233], and by the electromagnetic field modification of the particle in a medium, i.e., modification of the angular distribution of the virtual photons accompanying the particle in a medium [216,232].

The density effect impacts both the peak intensity and the angular divergence, setting an upper limit to these quantities, which does not change even with a further increase in electron acceleration energies. This effect is considerable for electrons with Lorentz factor $\gamma_e \geq \omega/\omega_p$, resulting in a saturation of the intensity and of the beam divergence. Below, we focus on the regime where the density effect is negligible, i.e., electron energies that satisfy $1 \ll \gamma_e \ll \omega/\omega_p$, leading to $\theta_{\text{ph}} \approx \gamma_e^{-1}$.

*3.4.2. The dispersion and angular distribution of PXR*

In the PXR emission process, two energy linewidths are generally of primary interest. The first is the full PXR energy linewidth, commonly also referred as the total PXR reflection linewidth, denoted by $\Delta\omega/\omega$. The second is the intrinsic PXR spectral linewidth, also called the spectral peak linewidth, represented by $\delta\omega/\omega$ (see **Figure 5**(b)). The full PXR energy linewidth, $\Delta\omega/\omega$, relates to the energy linewidth emitted from the entire reflection of the PXR across the entire angular opening of $\propto \gamma_e^{-1}$. In contrast, the intrinsic spectral linewidth, $\delta\omega/\omega$, represents the linewidth for a fixed (infinitesimal) emission angle. Due to the PXR's spatial dispersion, the intrinsic spectral linewidth $\delta\omega/\omega$ is typically much narrower than the full PXR linewidth $\Delta\omega/\omega$. This section details the characteristics of both the full PXR linewidth and the intrinsic linewidth, analyzing the PXR dispersion and angular distribution near each resonant (Bragg) emission point.

The specific resonant points governed by Bragg conditions (Eq. (3)). The dispersion relation, Eq. (3), enables extracting a relation of the angular spread and energy spread from the entire reflection region

$$\frac{\Delta\omega}{\omega} = \frac{\Delta\theta_B}{\tan\theta_B}, \qquad (6)$$

where $\Delta\theta_B$ is the angle deviation from the Bragg angle $\theta_B$, and $\Delta\omega$ is the energy deviation from the Bragg energy $\omega_B$ [155,162,234].

**Figure 5**(b) presents the PXR spatial dispersion around the resonant point $\Omega = 2\theta_B$. The PXR resonant energy ($\omega_B$) and resonant angle ($\theta_B$) are related by Eq. (3). The PXR emission has an angular opening of $\sim 2\gamma_e^{-1}$, where the peak intensity is located at angle $\theta_{\text{ph}} = \pm\gamma_e^{-1}$ relative to the Bragg angle. Thus, the full PXR linewidth is $\Delta\omega/\omega \sim \gamma_e^{-1}/\tan\theta_B$. This angular distribution of the PXR yield is often called PXR reflection [173].

In the resonant point $(\omega_B, \theta_B)$, the emission intensity is zero due to symmetry considerations: an electron penetrating the target material excites the material dipoles symmetrically, causing the dipole fields to cancel each other at the resonant point [153]. Therefore, the PXR geometry produces either a double lobe or a donut shape (**Figure 5**(a)) with a hole in the center. Other geometries break the symmetry and produce a PXR beam with a peak intensity exactly in the resonant point [153].

The PXR spatial dispersion is analogous to the transfer function of a crystal monochromator with the same parameters (i.e., the same material, Bragg plane, and angle). This property is advantageous for the PXR source since it allows excellent noise filtration schemes, analogous to the double monochromator scheme used in synchrotron facilities [177]. A further analysis and applications of this property are discussed in Section 5.3.

**Table 2**: Materials parameters for PXR yield calculations in Figure 5.

|  | Atomic number (Z) | Lattice Type | Unit cell dimensions [Å] | Absorption Length at 30 keV $L_{abs}$ [mm] | Radiation Length $X_0$ [mm] |
|---|---|---|---|---|---|
| **Graphite** | 6 | Hexagonal | $d_0 = 2.461$ $c = 6.708$ | 80.7 | 164 |
| **Aluminum** | 13 | FCC | 4.04 | 4.33 | 89.9 |
| **Silicon** | 14 | FCC | 5.43 | 3.77 | 94.8 |
| **Copper** | 29 | FCC | 3.61 | 0.11 | 14.7 |
| **Molybdenum** | 42 | BCC | 3.14 | 0.036 | 9.8 |
| **Tungsten** | 74 | BCC | 3.165 | 0.024 | 3.5 |

Under practical conditions, the observed width of the intrinsic PXR spectral linewidth, $\delta\omega/\omega$, at a specific observation angle $\Omega$ is primarily determined by the geometrical experimental angular resolution $\Delta\Omega_{geo}$. This angular resolution is influenced by factors such as the beam spot size on the crystalline target, the size of the X-ray detector, and the distance between them. The relationship between the PXR spectral linewidth and the angular resolution is given by $\Delta\omega/\omega = \Delta\Omega_{geo}/\tan\theta_d$, as shown experimentally in [162,234]. Section 5.2 provides a detailed discussion of the effective PXR linewidth, considering the experimental parameters such as the electron beam spot size, detector size, and the distance between the PXR crystal and the detector.

At ideal angular resolution ($\Delta\Omega_{geo} = 0$), the intrinsic PXR spectral linewidth $\delta\omega/\omega$ is determined by the number of crystallographic planes $N$ contributing to the PXR emission. This number is defined by the absorption length in the crystal (according to kinematical PXR theory) or by the extinction length (according to dynamical PXR theory). In cases for which these lengths are longer than the electron mean-free path, $N$ is instead determined by the mean-free path. The natural PXR linewidth relates to the number of crystallographic planes by $\delta\omega/\omega \propto N^{-1}$, which can be derived from both classical approaches and from the Heisenberg uncertainty principle [183]. An extremely narrow intrinsic PXR linewidth of $\delta\omega/\omega \sim 10^{-8}$ is achievable when relativistic particles moving in a channeling regime within a bent crystal emit a focused PXR beam [183]. However, for non-ideal angular resolution, $\Delta\Omega_{geo} \neq 0$, the intrinsic PXR linewidth, $\delta\omega/\omega$, becomes significantly broader, primarily influenced by the geometry of the PXR system, as discussed further in Section 5.2.

*3.4.3. The polarization of PXR*

**Figure 5**(a) shows the PXR angular shape and polarization for different polar emission angles. The PXR polarization is linear at every point of the PXR reflection. The polarization structures differ between the PXR emission in the forward hemisphere, backward hemisphere, and at the perpendicular direction to the incident particle beam. The polarization structure has a hyperbolic shape in the forward hemisphere and a parabolic shape in the backward hemisphere. In the exact backward direction, the parabolic shape becomes a radial polarization structure similar to Cherenkov radiation or transition radiation. The polarization structure in the PXR reflection emitted at the perpendicular angle to the particle beam has a specific shape [168], but most radiation is polarized in only one direction. The kinematical PXR theory of polarization is in good agreement with experimental realization [167,168,170].

*3.4.4. The main contributions to the radiation yield of PXR*

The PXR yield is provided in Eq. (4) and depends on the diffraction efficiency and the geometrical factor $f_\text{geo}$. The diffraction efficiency describes the PXR photons production per unit length. The geometrical factor, $f_\text{geo}$, captures the self-absorption of the PXR photons within the crystal during the emission process. Heavy materials have higher diffraction efficiency but are limited due to lower absorption length, which results in a smaller geometrical factor. Table 2 shows this tradeoff for different materials discussed in this review. For example, highly oriented pyrolytic graphite (HOPG) has a low atomic number and thus low diffraction efficiency, but it also has a higher absorption length and, thus, a higher geometrical factor.

*The diffraction efficiency* is calculated by the Fourier expansion of the electric susceptibility $\chi_g$ [84]:

$$\chi_g^2 = \frac{\lambda_x^4 r_e^2}{\pi^2 V_c^2} S_\text{hkl}^2 [(F_0(\boldsymbol{g}) + f_1 - Z)^2 + f_2^2], \tag{7}$$

where $\lambda_x$ is the emitted PXR wavelength, $r_e$ is the classical electron radius, $V_c$ is the volume of the crystal unit cell, $S_\text{hkl}$ is the structure factor, $Z$ is the atomic number, $\boldsymbol{g}$ is the reciprocal lattice wavevector, and $F_0(\boldsymbol{g}), f_1, f_2$ are the atomic form factors.

The term $F_0(\boldsymbol{g})$ is the momentum transfer efficiency of the beam, and can be described semi-analytically by the following expression [235]:

$$F_0(s) = \sum_{i=1}^{4} a_i \exp(-b_i s^2) + c, \tag{8}$$

where $s = \frac{\sin\theta_B}{\lambda_x} = \frac{1}{2d_{\text{hkl}}}$, and $a_i, b_i, c > 0$ are the Cromer-Mann coefficients [235,236]. Since $F_0(s)$ depends on $\exp(-b_i s^2) \propto \exp\left(-b_i \left(\frac{\sin\theta_B}{\lambda_x}\right)^2\right)$, the PXR yield decreases for higher PXR energies and larger PXR emission angles $\Omega$. Equivalently, the momentum transfer efficiency reduces for lower interplane distance $d_{\text{hkl}}$. This term limits the production of the PXR at high X-ray energies. To cope with this challenge, it is necessary to lower the Bragg angle. The atomic form factors $f_1$ and $f_2$ are the dispersion corrections, describing the behavior due to the bound inner-shell electrons; thus, they are independent of the wavevector **g** but depend only on the X-ray energy.

*The geometrical factor* is proportional to $f_{\text{geo}} \propto L_{\text{abs}} \propto 1/Z^4$ (section 4.2), whereas the diffraction efficiency is proportional to $\chi_g^2 \propto Z^2$, leading to a PXR yield dependence of $N_{\text{PXR}} \propto f_{\text{geo}} \chi_g^2 \propto 1/Z^2$. Therefore, lighter materials are preferable for producing more PXR photons. **Figure 5**(c) presents the PXR yield for various materials. Graphite (HOPG) is the lightest material examined (Z=6) and thus exhibits the highest yield. The typical values of PXR yield are $\sim 10^{-5} - 10^{-6}$ photons/ electron, and are calculated for optimal material thicknesses considering the absorption length. The jumps in the PXR yield (e.g., at ~8keV in Cu and at ~70keV in W) are due to the dispersion correction of the bound inner-shell electron cross-section (the $f_1$ and $f_2$ terms).

## 4. Recent developments toward practical applications

In this section, we present recent experimental and theoretical developments in PXR sources for increasing the flux to suit in-vivo bio-medical applications. Two parameters determine the PXR source flux – the yield (i.e., the average number of photons produced per single electron) and the electron source current (i.e., the number of electrons that pass through the target crystal per time unit). Even though the PXR yield is high relative to other electron-driven sources [126], the self-absorption of the emitted X-ray photons within the thick PXR crystal limits its yield [237]. Moreover, the thermal load on the PXR crystal restricts the maximal incident electron beam current [238].

We address these limitations in steps. In section 4.1, we present the progress in high-quality electron beam sources and their impact on the thermal load in the PXR target crystal. We discuss how state-of-the-art and next-generation electron sources can fit the thermal load requirement. In section 4.2, we review different PXR geometries that overcome the PXR photons' self-absorption limitation, enabling higher interaction lengths and higher spectral yield.

While the challenges and mitigation in this section related to the PXR emission, the insights and advancements discussed can be extended to other sources such as Smith-Purcell radiation, Cherenkov radiation, channeling radiation, and coherent bremsstrahlung [239].

## 4.1. Progress in PXR relying on high-quality electron beam sources

In recent years, progress in electron sources and acceleration structures has paved the way for high brightness, high-current electron sources, with practical applications for X-ray free-electron lasers (XFEL), ultrafast electron microscope (UEM), and ultrafast electron diffraction (UED) applications [143]. This progress leads to high electron source currents in compact acceleration structures. By using these novel high-current electron sources, the primary limiting factor transitions to the thermal load on the PXR crystal. Intuitively, the PXR source brightness increases with the number of electrons passing through the PXR crystal with the smallest possible spot size. However, the electron flux deposits energy in the crystal, leading to significant crystal heating and thermal vibrations that decrease the PXR yield. These considerations create a trade-off with a specific optimum. A recent quantitative analysis of this trade-off identified the optimal parameters, highlighting the prospects of a practical PXR source.

*4.1.1. The effect of heat load*

Relativistic electrons lose a small fraction of their kinetic energy when passing through a target. The energy loss goes partially into radiation emission (i.e., bremsstrahlung) and partially into heat. The heat from a single electron pulse is deposited in a volume determined by the electron beam spot size and the thickness of the PXR crystal. The thermal load causes crystal lattice vibrations, leading to phase mismatch between the atoms, and a degeneration of the constructive interference between the dipoles.

The PXR yield dependence on the crystal temperature is described by the Debye-Waller factor $e^{-2W}$ (Eq. (4)) [84]. Two distinct phenomena cause crystal lattice vibrations. The first is purely quantum mechanical and arises from the uncertainty principle. These vibrations are independent of temperature and occur even at absolute zero temperature, known as zero-point fluctuations. At finite temperatures, elastic waves (or phonons) are thermally excited in the crystal, increasing the amplitude of the vibrations. Those thermal vibrations cause PXR phase loss between the lattice dipoles, decreasing the PXR yield. This effect depends on the material-specific Debye temperature, $T_D$, the material temperature, $T$, and the $d$-spacing of the diffraction plane of interest, $d_{\text{hkl}}$.

The crystal thermal vibration mean square amplitude is given by [238]

$$u^2(T) = \frac{3\hbar^2}{4Mk_B T_D}\left[1 + 4\left(\frac{T}{T_D}\right)^2 \int_0^{T_D/T} \frac{y}{e^y - 1} dy\right], \qquad (9)$$

where $M$ is the material mass, and $k_B$ is the Boltzmann constant. The Debye-Waller term ($e^{-2W}$) is derived from the thermal vibration mean square amplitude ($u^2(T)$) and the reciprocal lattice vector ($\tau = 2\pi/d_{\text{hkl}}$), and equals to $e^{-2W} = \exp(-\tau^2 u^2(T))$. This relation with Eq. (9) leads to an exponential decrease in the PXR yield as the temperature increases. Due to this effect, there is an optimal electron current maximizing the PXR yield.

Previous PXR experiments were limited to average electron source current below 5 µA (Table 1). As stated in these experiments, using a larger electron charge per pulse caused damage to the PXR crystal [128]. However, by careful optimization of the electron source parameters, it becomes possible to increase the repetition rate of the electron source without damaging the PXR crystal. In essence, by refining the heat dissipation process, the optimal average electron source current varies depending on the PXR crystal material, falling within the range of ~500-3000 µA. This represents an increase of up to 2-3 orders of magnitude compared to the currents involved in prior X-ray imaging experiments (Table 1).

*4.1.2. The requirements from the electron beam source*

When increasing the electron source peak current, beam instabilities may emerge. This phenomenon is known as the beam blow-up (BBU) or the beam break instability [240]. It arises from the interaction between the electron beam and the cavity modes of the accelerating cells [241]. In this case, the electron beam is subject to density and velocity perturbation, increasing the beam emittance and energy spread. The higher the peak current, the more unstable the beam [242]; thus, to mitigate the electron BBU instabilities, a higher repetition rate with a lower peak current in each pulse is preferable. Indeed, next-generation X-ray FEL electron sources are designed to operate at a high repetition rate of 1 MHz [143].

It is important to highlight that a higher repetition-rate electron source is advantageous for the PXR scheme brightness due to the inverse relation between the optimal repetition rate and the electron beam spot size, leading to a smaller X-ray source spot size. For example, the optimal beam spot size for an electron source with a repetition rate of 1 MHz and pulse charge of ~1 nC is $\sigma_x \approx 40$ μm. Notably, state-of-the-art and next-generation electron sources fulfill the optimal requirements [142,243,244].

Additionally, it is worth noting that even if the electron beam quality has moderate degeneration, it would still meet the PXR source requirements. In contrast to the strict requirements of the X-ray FEL electron source, which must be satisfied for electron micro-bunching [3], the requirements for the PXR scheme are more relaxed, as discussed in section 3.4.

To further enhance the electron source peak current, a similar approach to the X-ray rotating anode tube can be used. X-ray tube machines experience similar heating challenges as the PXR crystal. The solution used in these machines is based on a rotating anode [84,245,246]. This method increases the effective heat dissipation area since the electron beam interacts with different positions of the target material. The PXR heat dissipation solution can use a similar approach. The main difference between the machines is that the target material for the PXR source should be modified by translation and not by rotation since a rotational change of the PXR crystal alters the X-ray emission direction. An additional crucial difference between the X-ray tube and the PXR source is that precision alignment is unnecessary with the X-ray tube but is critical for the PXR crystal. The alignment process can be similar to the double crystal monochromator scheme used in synchrotron facilities [247], where large crystals are available. These wafers can be translated much like a rotating anode so that the electron beam is concentrated near the outer edge of the wafer. This scheme can further increase the PXR flux, yet further study should explore the possible artifacts of a moving crystal target (such as blurring), as it has never been used before for PXR production.

### 4.2. Progress in PXR relying on material and geometry design

The emergence of new heterostructures and materials geometries, such as vdW materials, has led to precise and versatile methods of fabricating devices with atomic-scale accuracies. Hence, these materials have shown much promise for different technologies, including photodetectors, photocatalysis, photovoltaic devices, ultrafast photonic devices, and field-effect transistors [58]. By leveraging these advancements, the PXR crystal yield can be optimized, addressing challenges such as the self-absorption of PXR photons within the crystal. This section reviews recent breakthroughs, demonstrating the promising outcomes of utilizing such geometries.

*4.2.1. The challenge of X-ray self-absorption*

For a thick PXR crystal, the emitted PXR photons are self-absorbed within the crystal, limiting the contribution of all crystal layers to the PXR intensity (**Figure 6**(a)). This phenomenon is captured by the geometrical factor and sets an upper bound on the PXR yield. This limitation is especially significant for high-$Z$ materials with shorter absorption lengths.

Any X-ray beam gets attenuated during an interaction with a thick target material. The attenuation is caused due to several physical mechanisms but is mainly due to photoelectric absorption, Compton scattering, and elastic scattering [248]. The same phenomenon occurs for the emitted PXR photons within the crystal. Close to the crystal surface, the amount of PXR photons produced is linear with the material thickness. However, PXR photons that emit in deeper regions must traverse through the entire crystal, contributing significantly less than PXR photons produced at the surface of the crystal. Hence, the material absorption length limits the PXR yield.

The X-ray attenuation is exponential with an attenuation coefficient $\mu$, resulting in the following geometrical factor term expression [223]:

$$f_{\text{geo}} = L_{\text{abs}} \left|\frac{\hat{n} \cdot \widehat{\Omega}}{\hat{n} \cdot \hat{v}}\right| \left(1 - e^{-L/(L_{\text{abs}}|\hat{n} \cdot \widehat{\Omega}|)}\right), \tag{10}$$

where $L_{\text{abs}} = 1/\mu$ is the absorption length of the material, $\hat{n}$ is the normal to the crystal surface through which the electron beam transverse, $\widehat{\Omega}$ is the emission direction of the emitted PXR photon, $\hat{v}$ is the direction of the electron beam and $L$ is the crystal thickness. The attenuation coefficient is proportional to $\mu \propto \frac{Z^4}{\omega^3}$, depending on the X-ray energy, the material atomic number $Z$, and the material mass density; thus, heavier materials have much larger attenuation.

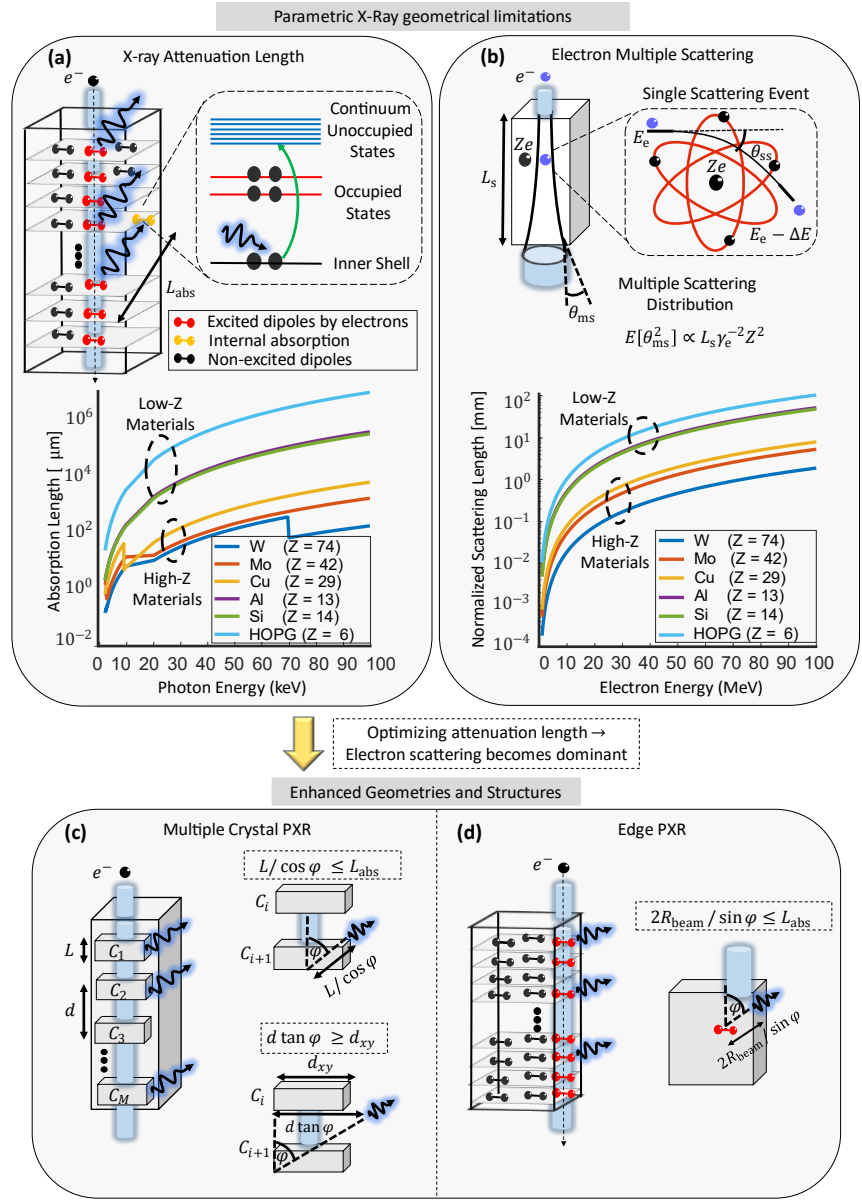

**Figure 6: PXR schemes for enhancing the PXR yield**. (a) X-ray attenuation length. The excited radiating dipoles produce PXR photons through all the crystal layers, yet the photons traversing the whole crystal attenuate. The absorption length is shown for different materials as a function of the PXR energy. High-$Z$ materials have shorter absorption lengths; thus, a smaller volume of the PXR crystal contributes to the emission. The absorption length is longer for higher PXR energies. (b) Electron beam multiple scattering. The electrons slightly deviate from their initial trajectory due to the electrostatic forces applied by the material atoms. The scattering length increases with higher electron energies. For heavier crystals, the scattering length is shorter. (c) Multiple crystals PXR scheme. The crystals are stacked upon each other with two fabrication conditions: 1) Each crystal should be thinner than the absorption length. 2) the distance between the crystals should be larger than the escape path of the emitted photon. (d) Edge PXR. The electron beam passes within the crystal in parallel to the crystal edge. To overcome the self-absorption of the emitted PXR photons, the beam spot size should be smaller than the attenuation length of the material. Another method for minimizing multiple scattering involves generating focused PXR through the channeling of positively charged particles, such as positrons, in a long, bent, thin crystal [183]. This approach is attractive as positrons have a significantly longer channeling length compared to electrons [249].

### 4.2.2. The challenge of electron beam scattering

When an electron passes through the PXR crystal, it gradually deviates from its initial trajectory due to the electrostatic forces applied by the material atoms. The electron scattering affects the PXR angular broadening, similar to the effect of the electron beam divergence. This scattering process has a random walk profile, for which the likelihood and the degree of an electron scattering is a probability function of the crystal thickness and the radiation length (i.e., the mean free path) [250].

In particular, the scattering angle is modeled with Gaussian probability with zero mean scattering and standard deviation. The following formula was empirically found to accurately capture the standard deviation of the scattering angle as a function of the electron energy $E_e$ and material type and thickness $L$ [251]:

$$\sigma_{\theta_{ms}} = \frac{13.6\text{MeV}}{E_e} \sqrt{\frac{L}{X_0}} \left(1 + 0.038 \ln\left(\frac{L}{X_0}\right)\right), \tag{11}$$

$X_0$ is the radiation length that depends on the material type. **Figure 6**(b) shows the electron scattering length for different materials and electron energies. Higher electron energies and lighter materials have lower scattering angles since $\sigma_{\theta_{ms}} \propto X_0^{-1/2} \gamma_e^{-1}$. The electrons' multiple scattering broadens the PXR angular shape, resulting in a higher PXR spectral linewidth (Eq. (6)).

Several methods were developed to evaluate the PXR angular broadening, as well as standard Monte Carlo numerical simulations [161,252,253]. Here, we present the Potylitsyn method, which agrees well with the experimental results [165]. In this method, the Gaussian distribution of the electron scattering is convolved with the PXR angular shape $N(\theta_x, \theta_y)$ (Eq. (5)):

$$\tilde{N}(\theta_x, \theta_y) = \frac{1}{2\pi \sigma_{\theta_{ms}}^2} \iint_{-\infty}^{\infty} d\phi_x d\phi_y \, N(\theta_x - \phi_x, \theta_y - \phi_y) \exp\left\{-\frac{(\phi_x^2 + \phi_y^2)}{2\sigma_{\theta_{ms}}^2}\right\}, \tag{12}$$

The electron multiple scattering leads to a spatial shape and dispersion broadening, which decreases the PXR source brightness, as discussed next.

*4.2.3. Overcoming the challenges by optimizing the crystal geometry*

We review two PXR schemes to cope with the limitation of self-absorption of the PXR photons within the crystal. In these geometrical schemes, instead of self-absorption, the limiting factor is the electron beam scattering (**Figure 6**(b)). The electron beam scattering leads to the PXR angular emission broadening and limits the number of emitted photons that hit the detector within the angular aperture. Assuming the self-absorption phenomenon is negligible, the optimal PXR crystal thickness is $L \approx 0.1 X_0$. Above this crystal thickness, the PXR flux gain becomes small, and the source brightness decreases. This optimal material thickness is larger by up to an order of magnitude than the absorption length of heavy materials. Since the X-ray attenuation coefficient is higher for lower X-ray energies, these schemes have a considerable gain for lower X-ray energies.

The first scheme is a stacked multiple crystals structure (**Figure 6**(c)), and the second is an edge PXR structure (**Figure 6**(d)). In the first scheme, two conditions should be fulfilled: (1) The thickness of each crystal should be thinner than the absorption length. (2) The distance between the crystals should be large enough for the emitted photons to not go through the adjacent crystal.

The "edge PXR" structure, which is also called "grazing PXR" or extremely asymmetric diffraction (EAD) PXR [145–147], is based upon transmission of the electron beam within the crystal, parallel to the crystal edge surface. In this structure, the electron spot size should be shorter than the absorption length for the emitted PXR photon traverse a shorter distance than the absorption length. This structure has been examined experimentally for silicon crystal, where a PXR yield gain by a factor of 5 was reported, which fits well with the theoretically expected gain [144].

### 4.2.4. Resulting optimal X-ray flux

**Figure 7** shows the PXR photon rate comparison between a standard PXR scheme and enhanced PXR schemes for different PXR materials. The X-ray spectrum is divided into the target applications: X-ray crystallography (<15keV), mammography (10-25keV), chest and head radiography (40-50keV), and abdomen and pelvis radiography (50-70keV). The dashed line represents the photon rate necessary for in-vivo imaging. The target's angular aperture used for flux derivation is $\theta_D \sim 3\gamma_e^{-1}$. The gain is considerable for lower X-ray energies due to the higher self-attenuation in this region. For higher X-ray energies, the flux decreases due to lower diffraction efficiency. Overall, the PXR flux for the different PXR crystals is adequate for practical applications.

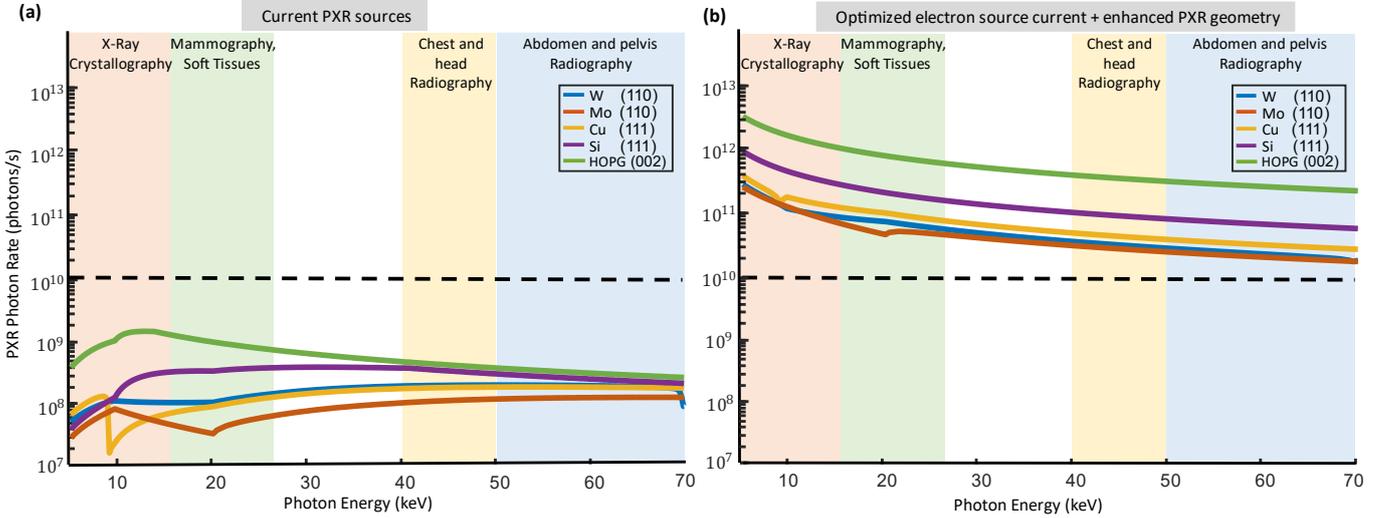

**Figure 7: Optimized PXR photon rate for optimized PXR geometries and electron source currents**. Photon rate comparison between a classic PXR scheme **(a)** and enhanced PXR schemes **(b)** for different materials, assuming an optimal electron source current and optimized geometries. The assumptions are: 1) the maximal interaction length is 10 mm (in both schemes). 2) The incident electron energy is 60 MeV. 3) The detector's angular aperture is $\theta_D = 3\gamma_e^{-1}$. 4) The material thickness in the regular PXR scheme is the material absorption length, while in the enhanced PXR geometry is $0.1X_0$, where $X_0$ is the material radiation length. The spectrum is split into regions for different applications. The dashed line marks the photon rate needed for practical applications. The enhanced PXR geometries have a more significant gain for lower PXR energies since the absorption length is shorter in these regions.

Such PXR schemes are limited by several challenges. In the multiple PXR crystals scheme, the final image may have a blurring artifact due to the many beams' emissions from each sub-crystal. Image processing techniques can reduce this artifact [64]. Moreover, the multiple crystals' alignment relative to the electron beam should be the same, which may be experimentally challenging. In the edge PXR scheme, the necessity for a precise alignment between the electron beam and the PXR crystal edge can face a challenge, yet earlier experiments demonstrated this [144]. Despite the PXR source flux growth, the PXR source signal-to-noise ratio remains the same between the standard and the enhanced schemes since both PXR and bremsstrahlung increase linearly with the material thickness.

It is important to highlight that recent experimental setups employ a wedge-shaped crystal plate as the PXR radiator [128,214]. In the case of a rectangle edge-shape, PXR beams emitted from the front and the side surfaces have different refraction properties. The superposition of these beams strongly disturbs phase-contrast imaging. Therefore, apart from enhancing flux, the primary motivation for the use of wedge-shaped crystal plates lies in suppressing multi-beam effects.

## 5. Roadmap toward a compact X-ray source

This section aims to present the applied aspects involved in developing compact sources of hard X-rays, with the ultimate goal of developing a viable source capable of important applications such as phase-contrast imaging. Key experimental considerations include electron beam quality, radiation safety, X-ray source geometry and dimensions, calibration process, and diagnostic systems. We specifically illustrate these considerations through the design of a compact PXR source, based on recent advances in the field. However, we note that these same considerations are not limited to PXR and apply to other compact X-ray sources such as ICS [29].

In section 5.1, we propose and analyze a design for a compact X-ray source. In section 5.2, we discuss the system performance and the emitted X-ray characteristics. In section 5.3, we discuss techniques for filtering noise from the X-ray source.

### 5.1. Design of the hard-X-ray source

**Figure 8** describes a compact source of hard X-rays based on the PXR mechanism. The electron source, based on a thermionic RF gun and a linear acceleration structure, produces the relativistic electron beam. Two Q-magnets focus the electron beam, one at the PXR crystal and the second before the electron beam dump. A double crystal scheme, based on a combination between a PXR crystal and a monochromator, produces a filtered PXR beam with a fixed exit location, as later in this section.

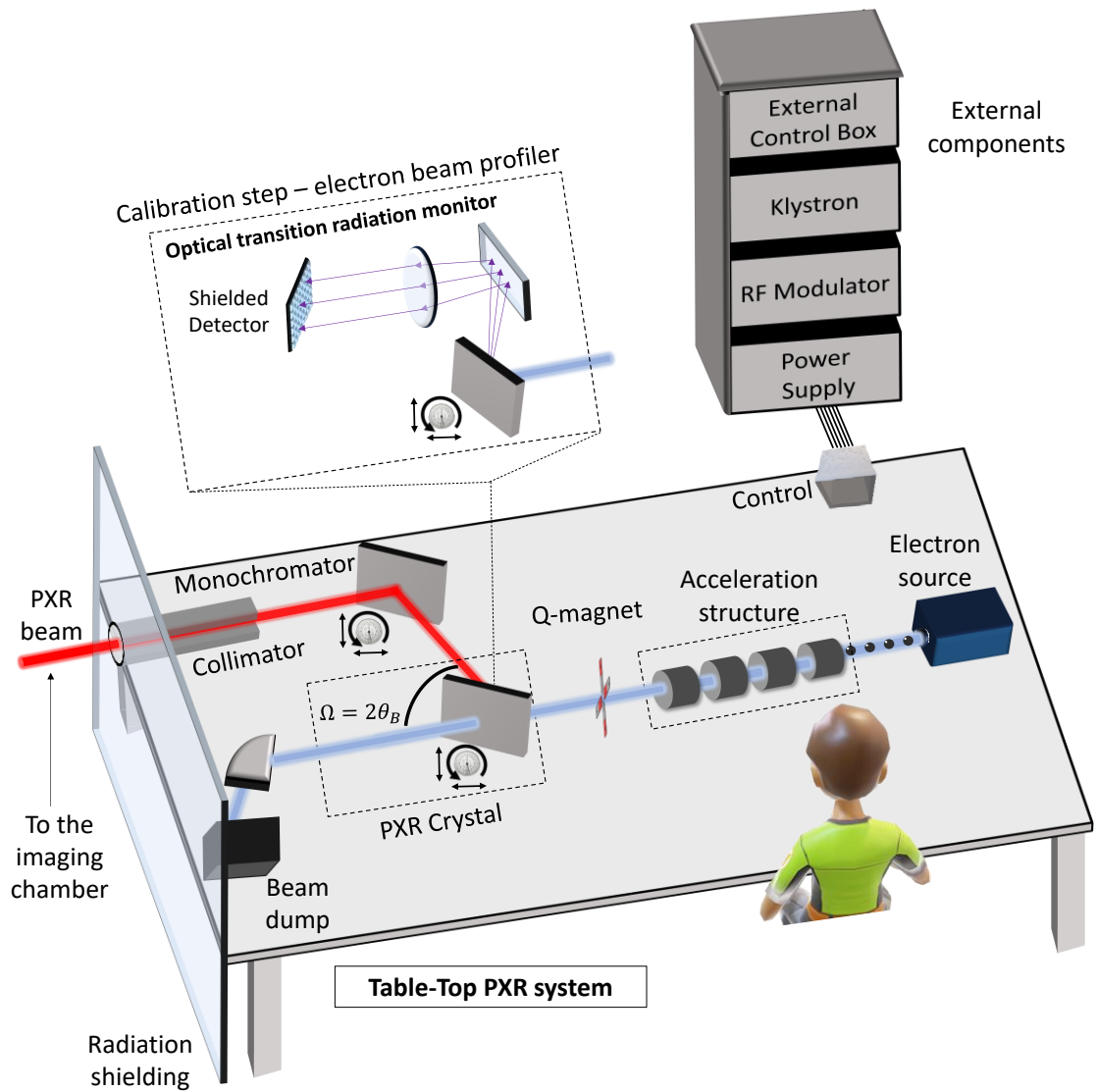

**Figure 8: Design for a compact system of parametric X-ray radiation (PXR).** The compact PXR system contains the following components: an electron source and a linear acceleration structure; a set of apertures and magnetic lenses that focus a collimated electron beam on the PXR crystal; the PXR crystal that produces the X-ray beam; a monochromator that filters the noise floor; additional quadruple magnet lens after the PXR crystal that focuses the electron beam to a beam dump, with a deceleration structure to reduce neutron production; an optical transition radiation (OTR) subsystem that monitors the electron beam position and width on the PXR crystal (observed in the backward geometry to avoid detection along the electron path and avoid the forward bremsstrahlung radiation). During the calibration stage, a control system optimizes the X-ray radiation by analyzing the optical signal and adjusting the electron beam position and the crystal displacement to position the beam correctly on the target. Additional power and RF modulation mechanisms are located outside the shielded PXR environment. A goniometer rotation mechanism (accuracy of ~0.01° [177]) and x-y displacement stages are used for the calibration process of both crystals. To optimize the X-ray beam quality, the monochromator crystal requires precise alignment relative to the PXR radiator, compensating for the slight difference between PXR frequency and Bragg frequency reflected by the monochromator (Eq. (3)) [134,222]. The estimated system size is roughly $3\times3$ m$^2$ [194].

The PXR beam exits through a collimator and an exit window. A power supply, RF modulator, a Klystron, and a control system feed the PXR system. The estimated dimension of the PXR source is ~3×3m², similarly to other tunable and compact X-ray machines, such as the inverse Compton scattering X-ray source [29]. In this design, the PXR crystal geometry can be based either on a regular or an advanced structure (section 4.2). The PXR crystal has an assembled cooler, and a heat sink to dissipate the heat from its edges.

The setup shown in **Figure 8** produces not only PXR but additional types of electromagnetic radiation, such as bremsstrahlung. Most radiation mechanisms are emitted in the forward direction. Bremsstrahlung radiation is produced by accelerated electrons in the different components of the systems, i.e., the electron gun, acceleration structures, collimators, the exit window, and the crystalline target. Additionally, the crystalline structure of the target generates coherent bremsstrahlung, while a wide-band forward-directed transition radiation is produced at the surfaces of both the exit window and the crystalline target.

Other types of radiation are emitted from the crystalline target at large angles relative to the beam axis. Among these are optical transition radiation, emitted at the mirror angle from the entrance surface of the target, and two distinct types of X-ray radiation besides PXR. The first type is isotropic characteristic X-ray radiation from the crystal atoms, characterized by fixed spectral peak energies. An additional type of radiation is produced during this process by the reflection of diffracted transition X-ray radiation at the Bragg frequency, which is coaxial with PXR reflection in the Bragg geometry [95].

A facility with an electron beam energy of up to approximately 100 MeV and a current of about 1 mA would support research and development applications for all these types of radiation, including PXR. These values are of particular relevance since they bring PXR to a regime where it dominates the overall emission and can be used for desired X-ray phase-imaging applications. Such a facility should be equipped with goniometer stages with angular precision of less than $0.1\gamma_e^{-1}$, X-ray detectors and spectrometers, as well as flexible tuning knobs to control electron beam parameters such as its angular spread and brightness.

The optical transition radiation (OTR) subsystem monitors the electron beam crossing with the PXR crystal. Generally, several mechanisms can accomplish this: (OTR) screen, YAG, wire scanner screen, and Cherenkov radiation [254]. Here we analyze the use of OTR, as it is broadly used in beam diagnostics in linear accelerators. Its linear intensity growth as a function of the beam current is a great advantage compared with fluorescent screens that are subject to saturation [255]. In addition, previous PXR experiments have used OTR for this purpose [152,202,215].

When considering high electron energy facilities, radiation safety is a central challenge to cope with due to the production of neutrons during the electron beam dump. The typical electron source energy required for a PXR source exceeds the neutron production threshold; thus, the PXR source must have a large and thick radiation shield to protect the operators and users. Several options can be employed to reduce the shielding requirements. The first option, proposed for ICS sources, is based on a deceleration structure before the electron beam dump. Using this technique allows fitting the source into a sea container. This option is presented in **Figure 8** and has been proposed previously also for a PXR source [137]. Another option is to use an electron beam energy below the neutron production threshold. Since the PXR emission energy does not depend on the incident electron energy, the PXR source scheme remains essentially the same. In this case, the PXR beam divergence would increase, yet it can be favorable for imaging applications due to the larger field of view. However, using electron source energies below ~10MeV for PXR imaging applications should be further researched.

### 5.2. System performance and X-ray source characteristics

In Section 3.4.2, we have seen that in an ideal PXR system, the intrinsic spectral linewidth is inversely proportional to the number of crystallographic planes, $\delta\omega/\omega \propto N^{-1}$. In the following section, we analyze the factors contributing to broadening this intrinsic PXR linewidth, $\delta\omega/\omega$, and conclude by determining the realistic linewidth achievable considering all these factors.

The intrinsic PXR spectral linewidth $\delta\omega/\omega$ is affected by several characteristics of the source, including the electron beam source quality, the PXR crystal material, and the experimental geometry [234]. These effects can be classified into three types of parameters: (1) Geometrical parameters, including the distance from the crystal to the detector $R_d$ and the detector collimation width $D_d$. (2) Crystal thickness and quality, especially its mosaicity, which represents the imperfection in the lattice translation throughout the crystal. (3) Electron beam quality, including its spot diameter $D_e$ and divergence $\Delta\theta_e$.

The electron source quality parameters, such as its energy spread and emittance, affect the performance of all high-brightness X-ray mechanisms [143]. However, in contrast with the other mechanisms, PXR in the ultra-relativistic regime is practically independent of the incident electron energy. Thus, the electron energy spread has a negligible impact on the intrinsic PXR linewidth $\delta\omega/\omega$. The linewidth still depends on the electron emittance, in addition to its strong dependence on the system geometry and the crystal Bragg plane and Bragg angle (Eq. (3)). The two subsections below elaborate on these considerations.

*5.2.1. Effects of geometry and electron beam quality on the X-ray linewidth*

When an incident electron impacts the crystal with a deviation angle $\Delta\theta_e$, both the Bragg angle ($\theta_B$) and the observation angle ($\Omega$) are shifted by the same amount, $\Delta\theta_e$. These parameters alter the PXR frequency, as can be captured by approximating Eq. (3):

$$\hbar\omega_{\text{PXR}} = \frac{2\pi\hbar c}{d_{hkl}} \frac{\sin(\theta_B + \Delta\theta_e)}{1 - \beta\sqrt{\epsilon}\cos(\Omega + \Delta\theta_e)} \approx \frac{2\pi\hbar c}{d_{hkl}} \frac{\sin(\theta_B + \Delta\theta_e)}{2\sin^2[(\Omega + \Delta\theta_e)/2]}, \quad (13)$$

which is valid for $\Delta\theta_e \ll 1$, $\Delta\theta_e \ll \theta_B$ and $\Omega \approx 2\theta_B$. This approximation helps to extract the intrinsic PXR spectral linewidth ($\delta\omega/\omega$) broadening and its dependence on two primary factors: the first is related to electron beam divergence and its multiple scattering captured by uncertainties in $\Delta\theta_e$, and the second arises from geometrical uncertainties in the observation angle $\Omega$. **Figure 9** illustrates the impact of these factors on the intrinsic PXR spectral linewidth ($\delta\omega/\omega$) broadening.

For the first factor, the uncertainty in $\Delta\theta_e$ (**Figure 9**(a)), the first-order derivative of the PXR energy with respect to $\Delta\theta_e$ is zero in the central observation angle (the Bragg angle $\Omega = 2\theta_B$):

$$\left.\frac{\partial(\hbar\omega_{\text{PXR}})}{\partial(\Delta\theta_e)}\right|_{\Delta\theta_e=0,\Omega=2\theta_B} = 0. \quad (14)$$

Therefore, Eq. (14) implies that the first-order effect of the electron beam divergence vanishes, and only second-order effects contribute, i.e., the effect of multiple scattering on the intrinsic PXR linewidth broadening is relatively small. Consequently, the second-order approximation for the intrinsic spectral linewidth broadening is $\delta\omega_{\text{div}}/\omega \approx \frac{1}{4}\frac{(\Delta\theta_e)^2}{\sin^2\theta_B}$, which is typically smaller than the other contributions for the intrinsic spectral linewidth broadening as discussed further below.

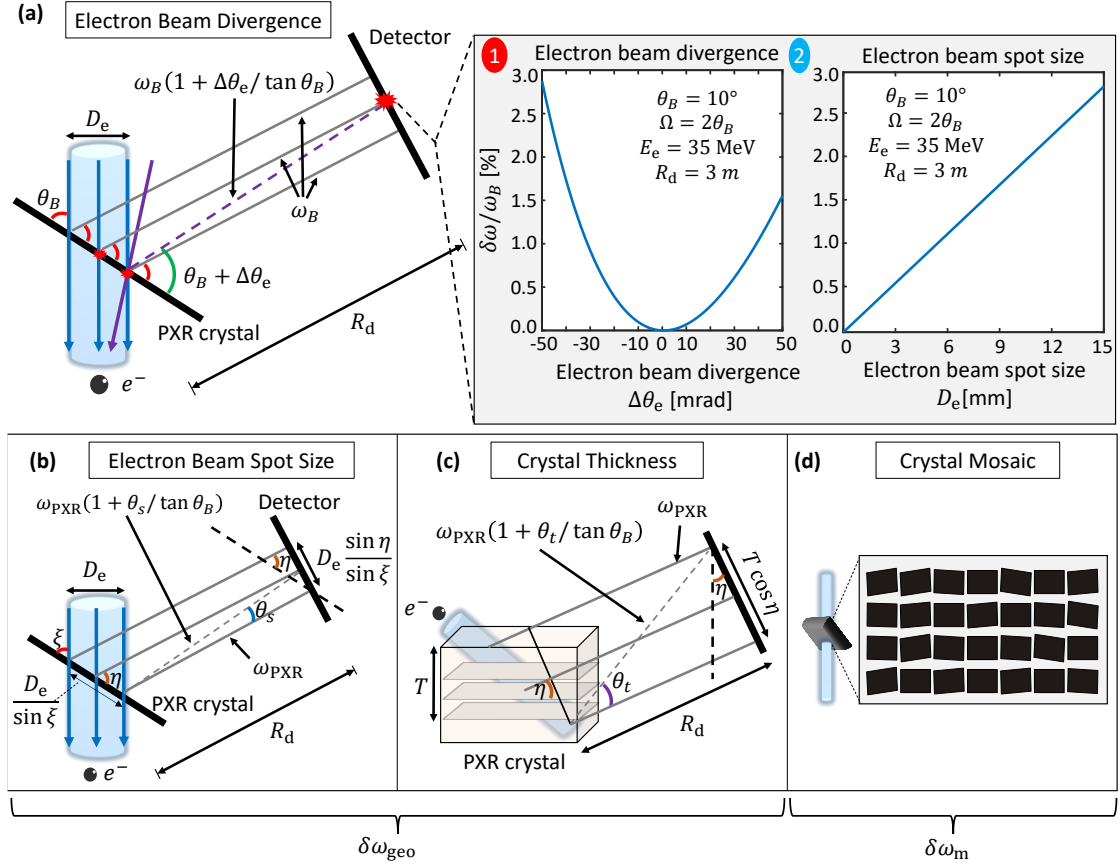

**Figure 9: The influence of experimental factors on the spectral X-ray linewidth $\delta\omega/\omega$.** The dependence of the PXR linewidth, $\delta\omega/\omega$, on the electron beam divergence (a) and spot size (b), crystal thickness (c), and crystal mosaicity (d). A collimated electron beam impacts the PXR crystal with an angle ξ relative to its surface (panel (a) assumes that the crystal surface and the Bragg plane coincide). The PXR material's dipoles radiate into an angle $\eta$ relative to the PXR crystal surface. The radiation arises from a confined volume in the PXR crystal, defined by the crystal thickness, electron beam trajectory, and its spot size. The impact of electron beam divergence and spot size on the PXR energy spread is shown in panels (a1) and (a2), respectively. The parameters used for producing the graphs are the electron beam energy of $E_e = 35$ MeV, the distance between the PXR crystal and the detector is $R_d = 3$ m, the Bragg angle is $\theta_B = 10°$, and the observation angle is $\Omega = 2\theta_B$. The uncertainty in the location of the radiating dipole results in geometrical linewidth broadening, as the emission energy is related to the emission observation angle (Eq. (3)).

The uncertainty in the observation angle Ω originates from the geometrical factors of the electron beam spot size (**Figure 9**(a2,b)), the thickness and absorption length of the crystal (**Figure 9**(c)), and the numerical aperture captured by each individual pixel in the detector (also called detector collimation), as follows:

$$\Delta\Omega_{\text{beamSpotSize}} = \frac{D_e}{R_d}\frac{\sin\eta}{\sin\xi}, \qquad (15)$$

$$\Delta\Omega_{\text{crystalThickness}} = \frac{\min(T, L_{\text{abs}}(\omega)\sin\eta)}{R_{\text{d}}}\cos\eta,$$

$$\Delta\Omega_{\text{detectorCollimation}} = \frac{D_{\text{d}}}{R_{\text{d}}},$$

where $0 \leq \xi \leq \pi/2$ is the angle between the target surface and the velocity vector of the electron beam, $0 \leq \eta \leq \pi/2$ is the angle between the target surface and the observation direction (See **Figure 9** for illustration), and $L_{\text{abs}}(\omega)$ represents the absorption length of the PXR photon with a frequency $\omega$ (Eq. (10)). The term $D_{\text{e}}\frac{\sin\eta}{\sin\xi}$ describes the electron beam spot size on the target surface in the observation plane, while the term $L_{\text{abs}}(\omega)\sin\eta\cos\eta$ represents the effective thickness of the target visible to the detector in the observation plane. The term $D_{\text{d}}$ represents the dimension of a *single pixel* of the detector. Therefore, combining all the terms in Eq. (15) leads to the following uncertainty in the observation angle:

$$\Delta\Omega_{\text{geo}} = \sqrt{\Delta\Omega^2_{\text{beamSpotSize}} + \Delta\Omega^2_{\text{crystalThickness}} + \Delta\Omega^2_{\text{detectorCollimation}}}. \qquad (16)$$

The linewidth broadening due to the geometrical angular uncertainty is [253]:

$$\frac{\delta\omega_{\text{geo}}}{\omega} = \frac{\Delta\Omega_{\text{geo}}}{\tan(\theta_B)}, \qquad (17)$$

where Eq. (17) applies both to Laue and Bragg geometries. In the limit of $\Delta\Omega_{\text{geo}} \to 0$, we get $\delta\omega/\omega \propto N^{-1}$, with $N$ the number of crystallographic planes (see Section 3.4).

The first two terms in Eq. (15) – the electron beam size ($\Delta\Omega_{\text{beamSpotSize}}$) and the crystal thickness ($\Delta\Omega_{\text{crystalThickness}}$) – are typically lower than $10^{-3}$ in most state-of-the-art PXR setups. In addition, in modern imaging applications that use high-resolution detectors, the ratio between the detector's dimension, $D_\Omega$, to the distance between the PXR crystal and the detector, $R_d$ (i.e., the detector's collimation term in Eq. (15)), is typically much smaller than the other two terms. Hence the detector's collimation has a minimal impact on the broadening of the intrinsic PXR spectral linewidth $\delta\omega/\omega$. Considering all the geometrical terms in Eq. (15), the typical intrinsic PXR linewidth is on the order of $\delta\omega/\omega \sim 1\%$. Moreover, it is important to note that the broadening of the intrinsic linewidth depends on $\cot\theta_B$ (Eq. (17)). Thus, since higher PXR energies require lower Bragg angles (Eq. (3)), it is preferable to use Bragg planes with higher Miller indices (i.e., smaller interplane distances) to maintain higher Bragg angles and minimize the intrinsic PXR spectral linewidth.

*5.2.2. Effect of crystal mosaicity on the X-ray linewidth*

The crystal mosaicity is an additional parameter affecting the intrinsic PXR spectral linewidth (**Figure 9**(d)). Mosaicism is the degree of imperfection in the lattice translation throughout the crystal [84]. Macroscopic crystals are often imperfect and composed of small perfect blocks with a distribution of orientations around some average value. Since each mosaic block emits a PXR beam with a slightly different orientation and angle, the PXR beam is spatially broadened, leading to a PXR spectral linewidth broadening. Typically, the mosaic blocks have orientations distributed over an angular range between 0.01° and 0.1° [84]. Graphite (HOPG), which has a high PXR yield, suffers from high mosaicity with an angular range of 0.4° [176]. The total broadening of the intrinsic linewidth, accounting for the geometrical factor, electron beam divergence, and the crystal mosaicity is given by [253]

$$\frac{\delta\omega}{\omega_B} = \frac{\sqrt{(\Delta\Omega_{\text{geo}})^2 + (\Delta\theta_m)^2}}{\tan\theta_B} + \frac{1}{4}\frac{(\Delta\theta_e)^2}{\sin^2\theta_B}, \qquad (18)$$

where $\Delta\theta_m$ account for the crystal mosaicity. Note that mechanical tensions on the crystals can increase $\Delta\theta_m$, and can also be used to induce intentional variation in the lattice translation throughout the crystal to facilitate X-ray focusing [71,72].

### 5.3. Optimization of the X-ray source signal-to-noise-ratio

The radiation emitted from the source includes the desired PXR and competing mechanisms such as bremsstrahlung and transition radiation. Achieving a high X-ray beam quality requires filtering these competing mechanisms, as they act as a broadband noise that diminishes the brightness of PXR [256]. One strategy to mitigate the noise involves optimizing the PXR experimental parameters by enlarging the PXR emission angle, thereby reducing the intensity of bremsstrahlung and transition radiation in the detector plane. While bremsstrahlung and transition radiation are emitted in the forward direction within a narrow cone of $\gamma_e^{-1}$, PXR can be emitted at a large emission angle of $\Omega \gg \gamma_e^{-1}$. By enlarging the PXR emission angle, bremsstrahlung and transition radiation become less intense in the detector plane [138].

While larger PXR angles can help reduce the noise floor, additional noise suppression is essential, especially at higher X-ray energies. A common approach involves using a double-crystal system (illustrated in **Figure 8**) that acts as a bandpass filter, with its passband energy range aligned with the PXR spatial dispersion [177,257,258]. In this double-crystal system, the PXR crystal and monochromator are arranged in a non-dispersive configuration (**Figure 10**(a)), similar to those used for filtration in synchrotron facilities [259] [260]. Unlike these conventional monochromator designs, where the X-ray beam passes through two crystals that are held parallel, in the PXR scheme the electron beam only interacts with one crystal to produce an X-ray beam (i.e., the PXR radiator). This X-ray beam then impinges the second crystal, which is aligned to reflect only the PXR energy, acting as a crystal monochromator. This process ensures that the X-ray beam exits in the same direction as the incoming electron beam (as shown in **Figure 8** and **Figure 10**(a)), allowing for consistent beam extraction without moving the entire PXR source or the target, thus making it highly advantageous for stable X-ray production [178]. In other words, this setup solves the challenge of maintaining a fixed output port for X-ray extraction, independent of the choice of PXR emission angle and the crystal angle [177].

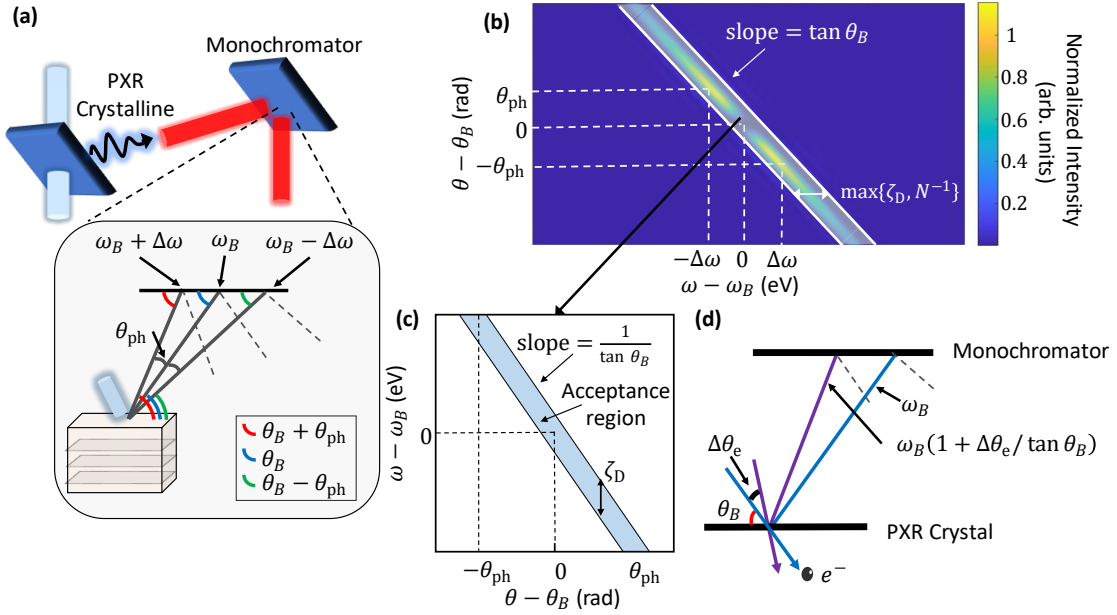

**Figure 10: Crystal monochromator for bandpass filtering of PXR. (a)** Suppressing the noise floor with a monochromator. The PXR beam impacts the monochromator crystal in the Bragg geometry with a central angle $\theta_B$ relative to the Bragg plane. Due to the PXR spatial dispersion shape, the corresponding frequencies for the emitted angles $[\theta_B - \theta_{\text{ph}}, \theta_B + \theta_{\text{ph}}]$ are $[\omega_B - \Delta\omega, \omega_B + \Delta\omega]$, respectively, where $\Delta\omega/\omega_B = \theta_{\text{ph}}/\tan\theta_B$ (Eq.(6)). If the monochromator has the same parameters as the PXR crystal (i.e., the same material, Bragg angle, Bragg plane), its transfer function overlaps with the incoming beam's spatial dispersion. The non-dispersive arrangement between the PXR crystal and the monochromator preserves the PXR beam exit location under equal rotations of the PXR crystal and of the monochromator, removing the necessity to rotate the whole PXR machine or the target sample during the PXR energy tuning process. **(b)** The PXR spatial dispersion from a single electron. The intensity is zero in the center frequency $\omega_B$ due to the PXR symmetry properties. The maximum intensity location is at the emission angles $\theta_B + \theta_{\text{ph}}$ and $\theta_B - \theta_{\text{ph}}$. When fixing the emission angle, the intrinsic PXR linewidth $\delta\omega/\omega$ scales as $N^{-1}$, for $\gamma_e \gg 1$ or else the electron mean-free-path limits the linewidth too. **(c)** The DuMond diagram of a crystal monochromator. The accepted linewidth for a fixed incident angle is the Darwin width $\zeta_D$. The slope of the acceptance region is $1/\tan\theta_B$, and the incident beam divergence is $\theta_{\text{ph}}$; thus, the emitted spectrum has a linewidth of $\Delta\omega/\omega_B = \theta_{\text{ph}}/\tan\theta_B$. The monochromator transfer function contains the PXR's spatial dispersion. **(d)** Illustration of the non-dispersive crystal arrangement in symmetric Bragg geometry. The central axis of the electron beam (blue line) is reflected by the first crystal (the PXR crystal). Then, the X-ray beam would be reflected by the second crystal (the monochromator crystal) and would emerge parallel to the axis of the incident electron beam. An electron incident at an angle $\Delta\theta_e$ relative to the central ray (purple line) will be Bragg reflected with an angle $\theta_B + \Delta\theta_e$ relative to the Bragg plane and with the X-ray energy of $\omega_B(1 + \Delta\theta_e/\tan\theta_B)$. Consequently, the emitted X-ray beam will emerge parallel to the deviated incident electron.

Generally, the rocking curve of a monochromator is very narrow, as described by the DuMond diagram (**Figure 10**(c)) [217]. The DuMond diagram describes the monochromator transfer function and acceptance region as a function of the incident X-ray energy ($\omega_B$) and the angle relative to the Bragg plane ($\theta_B$). An incoming X-ray beam that satisfies the Bragg condition will be reflected from the monochromator at the same angle as the incident beam angle. However, an incident beam that is slightly off the Bragg condition will be attenuated by the monochromator. Thus, the diffracted intensity of a polychromatic X-ray beam from a monochromator can drop by up to four orders of magnitude, limiting the flux considerably.

The effectiveness of the double-crystal system for the PXR filtration is possible due to the PXR spatial dispersion (**Figure 10**(b)). As discussed in Section 3.4.2, the spatial dispersion of PXR and the transfer function of a crystal monochromator with matching parameters to the PXR crystals almost perfectly align, allowing for an efficient filtration process (**Figure 10**(b) and **Figure 10**(c)). However, according to Eq. (3), there is a slight difference between the PXR frequency and the Bragg frequency reflected by the monochromator [134,222], requiring a small compensation between the two. This difference can be compensated with precise adjustments and fine-tuning of the crystal alignment., i.e., the monochromator crystal is tilted relative to the PXR radiator with precision on the order of ∼0.01° [177]. This level of precision is possible for example by utilizing nano-piezo goniometers [261]. Thus, by carefully optimizing the crystal arrangement, the PXR signal passes through the monochromator with minimal attenuation, while the noise floor is largely filtered out [138,177]. This selective reflection enhances the overall efficiency of the PXR-based X-ray source.

# 6. Comparison of the leading compact hard X-ray sources

The leading mechanisms for hard-X-ray generation in compact scales are inverse Compton scattering (ICS), characteristic radiation produced from an X-ray tube, and PXR. In this section, we compare these mechanisms using the metrics of spectral yield, flux, brightness, and practical application suitability. Specifically, for each mechanism, we analyze the energy tunability, source dimensions, aspects of radiation safety, operational simplicity, and requirements of the active components (i.e., the electron beam and laser sources). Table 3 summarizes this comparison.

**Table 3:** Comparison between parametric X-ray (PXR), inverse Compton scattering (ICS), and characteristic radiation. The red, orange, and green colors represent disadvantage, slight advantage, and advantage properties, respectively.

| | Parameter | Characteristic X-ray | Inverse Compton Scattering | Parametric X-ray |
|---|---|---|---|---|
| **X-ray beam** | Energy tunability | Limited by inner shell transition energies | Electron energy, laser wavelength | Crystal rotation, Bragg plane, PXR material |
| | Energy linewidth | <0.1% | ~1% | $\theta_D / \tan\theta_B$ (Eq. (6)) |
| | Emission cone | $4\pi$ sr | $0.1\gamma_e^{-1}$ | $\gamma_e^{-1}$ |
| | Spatial dispersion shape | - | Parabolic (**Figure 11**(a)) | Chirp (**Figure 5**), overlaps with a monochromator transfer function |
| | Spectral yield $\left[\frac{\text{photons}}{\text{electron}}\right]$ | $10^{-7} - 10^{-6}$ | $10^{-4}$ | $10^{-5} - 10^{-4}$ |
| | Average Brightness $\left[\frac{\text{photons}}{\text{s mm}^2 \text{ mrad}^2}\right]$ | $10^{10} - 10^{11}$ | $10^{13} - 10^{14}$ | $10^{11} - 10^{12}$ |
| **Electron source** | Energy | 100 keV | 8 MeV – 50 MeV | 50 MeV |
| | Normalized emittance | No impact | Impacts the X-ray source emittance and energy linewidth | Less strict requirements than ICS |
| | Energy spread | No impact | Impacts the X-ray energy linewidth | Negligible impact |
| **Machine** | Dimensions | Mobile | Table-top | Table-top |
| | Operational Simplicity | Simple | Spatial and temporal alignment between the laser and electron beams | Spatial alignment only between the electron beam and PXR crystal |
| | Neutron radiation safety | No requirements | Neutron shielding, an electron beam deceleration structure | |

Figure 11 compares the flux and brightness between the PXR, ICS, and characteristic X-ray sources. The PXR source flux is the highest, particularly for lower X-ray energies, i.e., it may serve as a promising imaging technique for applications in this spectrum range, such as mammography. However, the PXR flux decreases for higher X-ray energies due to lower diffraction yield. The characteristic lines produced from a rotating anode have a high flux due to the usage of high electron source currents. However, the X-ray flux emitted from the liquid-jet anode is much lower since the electron source average current is significantly lower.

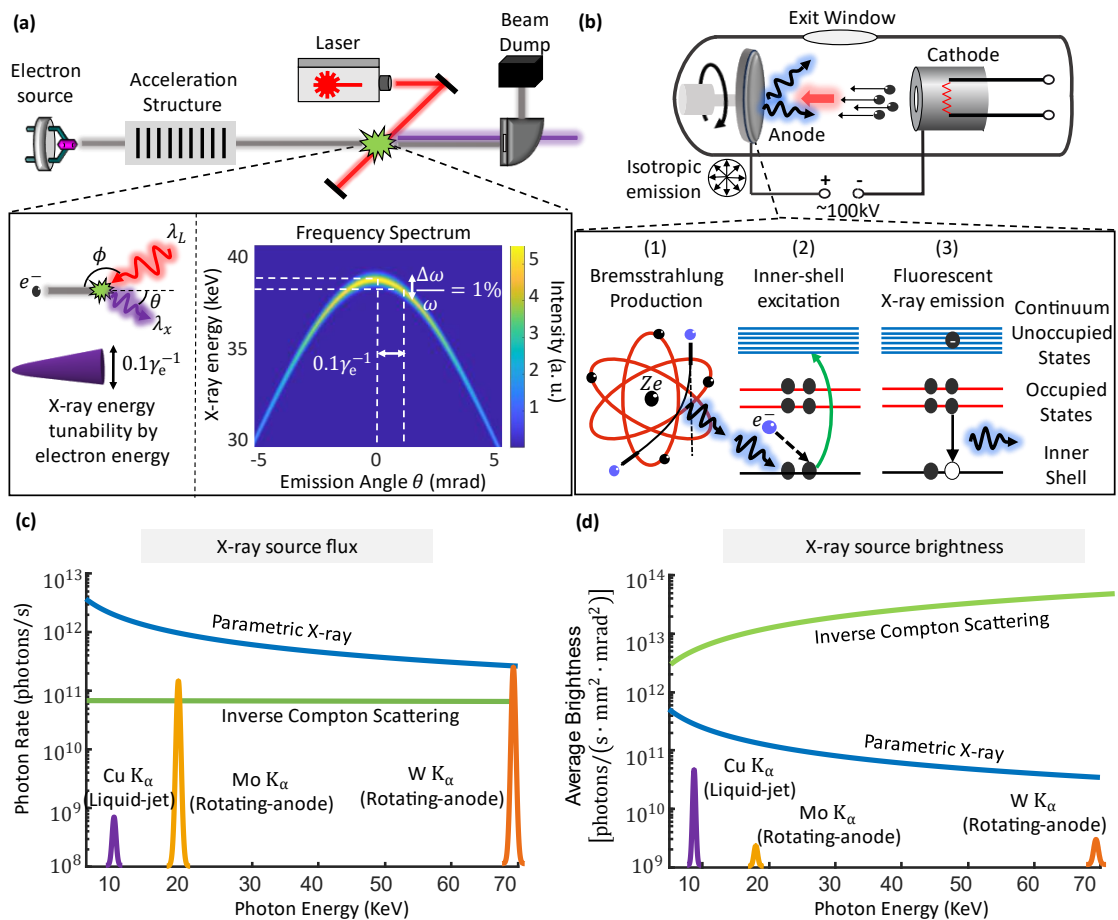

**Figure 11: Comparison between Parametric X-ray (PXR) to inverse Compton scattering (ICS) and characteristic radiation. (a)** Inverse Compton scattering X-ray source scheme. A relativistic electron beam collides head-on with a laser pulse, upconverting the laser photon energy to an X-ray photon. Within a narrow emission cone $\sim 0.1\gamma_e^{-1}$ in the forward direction, the X-ray beam energy linewidth is 1%. The spatial dispersion of the ICS beam is parabolic. **(b)** Characteristic radiation production from a rotating anode X-ray tube. Electrons are emitted by thermionic emission from a filament cathode and accelerated to < 100kV. They hit a rotating anode target, producing isotropic bremsstrahlung and characteristic radiation. **(c)** Flux comparison between PXR, ICS, and characteristic radiation. For the flux derivation, we assume

the target angular aperture is 15mrad. **(d)** Brightness comparison between PXR, ICS, and characteristic radiation: The characteristic brightness is calculated for the K-line linewidth, the PXR brightness is calculated within the linewidth defined by the emission cone, and the ICS brightness is calculated within 1% linewidth. Table 4 summarizes the experimental parameters used for producing the graphs.

When comparing the X-ray sources' brightness, both ICS and liquid-jet X-ray tubes gain a significant advantage. Both sources use high-brightness electron sources. The ICS brightness increases with the X-ray energy since the ICS beam divergence decreases proportionally to the inverse of the electron energy ($\gamma_e^{-1}$). In contrast, the PXR angular divergence is independent of the emitted X-ray energy but only of the electron beam energy. However, for imaging applications, a large field of view is advantageous; thus, the flux is a more representative metric compared with the brightness.

**Table 4:** Parameters used for flux and brightness calculation.

| | Parameter | Characteristic Line W $K_\alpha$ | Characteristic Line Mo $K_\alpha$ | Characteristic Line Cu $K_\alpha$ | Inverse Compton Scattering | Parametric X-ray |
|---|---|---|---|---|---|---|
| **Material** | Target Material | Tungsten Rotating anode | Molybdenum Rotating anode | Copper Liquid-jet | - | HOPG \ Diamond |
| **Electron source** | Electron energy | 100keV | 100keV | 100keV | 8-50 MeV | 50MeV |
| | Average electron current | 500mA | 500mA | 2mA | 10μA | 1mA |
| | Electron beam spot size | 1mm | 1mm | 10μm | 2μm | 40μm |
| | Average current density (mA/mm$^2$) | 500 | 500 | 20000 | 2500 | 600 |
| **Laser source** | Wavelength | - | - | - | 515 nm | - |
| | Pulse energy | - | - | - | 10 mJ | - |
| | Beam waist | - | - | - | 5 μm | - |
| | Repetition rate | - | - | - | 100 KHz | - |

## 6.1. Comparison with inverse Compton scattering (ICS)

ICS is the up-conversion process of a low-energy laser photon to a high-energy X-ray photon by scattering from a relativistic electron. **Figure 11**(a) shows the interaction scheme with a near head-on collision between the laser and electron beams. The scattered X-rays emerge in the same direction as the electrons. The physical mechanism of ICS is nearly identical to spontaneous synchrotron emission in a static magnetic undulator as used at traditional synchrotron facilities. However, due to the much shorter micro-meter laser wavelength, relative to the centimeter-period undulator wavelength, the required electron energies to produce hard X-ray photons are orders of magnitude lower than in the large synchrotrons [30].

### 6.1.1. The spatial and angular distribution of ICS

The up-conversion ratio for low laser intensity and on-axis emission from a head-on collision is given by [30,31]

$$\lambda_x = \frac{\lambda_L}{4\gamma_e^2}\left(1 + \gamma_e^2\theta^2 + \frac{a_0^2}{2}\right), \tag{19}$$

where $\theta$ is the X-ray photon emission angle relative to the electron beam direction, $a_0 = \frac{eE_0\lambda_L}{2\pi m_e c^2}$ is the dimensionless vector potential of the laser field, $\lambda_L$ is the laser wavelength and $\lambda_x$ is the emitted X-ray wavelength. The dimensionless vector potential value should be well below unity and typically $a_0 \leq 0.1$ (i.e., the linear ICS approximation) to avoid harmonic powers, and distortion of the energy linewidth [30].

**Figure 11**(a) shows the ICS parabolic spatial dispersion [30]. The up-conversion ratio (Eq. (19)) implies that all photons emitted within a narrow cone of $\sim 0.1\gamma_e^{-1}$ have an energy linewidth of 1%. While the low beam angular divergence is advantageous for high-brightness applications, it is a disadvantage for imaging applications that require

a large field of view since it requires a long distance to the target. In addition, the ICS parabolic spatial dispersion does not overlap with the crystal monochromator transfer function, as opposed to the chirp shape of PXR spatial dispersion. Therefore, without appropriate treatment, the ICS beam would be significantly attenuated by the monochromator. A scheme based on a Kirkpatrick–Baez (KB) mirror combined with a double crystal monochromator for focusing and filtering the beam was proposed to cope with this challenge, resulting in a 60% flux reduction [29].

*6.1.2. The yield of ICS*

The total number of ICS photons produced over all angles and frequencies is determined by the cross-section between the electron beam and the laser photons:

$$N_x = \frac{N_e N_L \sigma_T}{2\pi(\sigma_L^2 + \sigma_e^2)}, \qquad (20)$$

where $\sigma_T$ is the Thomson cross-section, $N_e$ is the total number of electrons, $N_L$ is the total number of photons in the laser beam, and $\sigma_L$ and $\sigma_e$ are the beam spot size at the interaction point of the laser and electron beam, respectively. For an ICS scheme with a laser wavelength of $\lambda_L = 515$nm, laser pulse energy of 10mJ, a laser beam waist of 5μm, and $a_0 = 0.1$, the ICS yield is $\sim 10^{-3}$ photons/electron, accounting for the ICS emissions in all directions and all frequencies. However, due to its spatial dispersion, the ICS spectral yield, accounting only for photons emitted at 1% linewidth, is more than an order below [29]. Thus, the ICS spectral yield is comparable with the PXR yield from a HOPG \ diamond crystal with an optimal geometry.

*6.1.3. The challenges with ICS*

The ICS scheme requires geometrical and temporal synchronization between high-quality electron and laser beams. For a scattering process such as ICS, the highest flux is produced by squeezing the electron and laser beams into a small spot size with a short duration. In this scheme, the electron source emittance determines the emitted X-ray beam emittance; thus, the electron source emittance must be low, typically a few orders of magnitude lower than the requirement for the PXR source (section 5.2). Moreover, since the up-conversion ratio is directly proportional to the laser photon energy and the electron beam energy (Eq. (19)), the ICS source must use a low laser linewidth and a low electron beam energy spread to produce a low linewidth X-ray beam [30].

## 6.2. Comparison with characteristic radiation

Due to its simplicity, characteristic radiation produced from an X-ray tube is the most widespread emission mechanism when a monoenergetic X-ray beam in a laboratory-scale facility is necessary. This emission occurs when an electron is accelerated from a hot cathode and impacts a target anode (**Figure 11**(b)). The characteristic X-ray photon emission includes inner-shell electron photoionization followed by fluorescence emission. If the incident electron kinetic energy is larger than the inner-shell binding energy, it knocks out the inner-shell electron and produces a vacancy. The ionization process can occur either by a direct electron impact or a bremsstrahlung photon. Typically, the inner-shell ionization cross-section by a direct electron impact is two orders of magnitude higher than the bremsstrahlung inner-shell ionization cross-section [262]. A comparison of characteristic X-ray radiation yield and PXR yield excited by relativistic electrons in the Si crystal can be found in [95].

### 6.2.1. The yield of characteristic radiation

Following ionization, an electron from an outer shell fills the vacancy in the ionized inner shell. In this process, the energy between the two bound states is emitted either in a radiative way with a characteristic X-ray photon (i.e., a fluorescence process) or by a non-radiative process. In the non-radiative process, another bound electron is emitted from the atom, a process known as Auger electron emission [262]. The fluorescence yield, $Y_f(Z)$, describes the probability of fluorescence emission as a function of the material's atomic number and can be approximated by [263]

$$Y_f(Z) = Z^4/(Z^4 + a), \tag{21}$$

where $a = 1.12 \times 10^6$. Experimental values for the fluorescence yield can be found in online databases [264]. The fluorescence yield increases for higher Z materials; thus, high-Z materials produce more intense characteristic lines.

The total number of emitted characteristic X-rays for the case of direct impact by a single incident electron is defined by the product of the ionization cross-section and the probability for fluorescence emission:

$$N_{\text{chr}}^{(t)} = \sigma_K n_a L(\omega_c^-) Y_f(Z), \tag{22}$$

where $\sigma_K$ is the cross-section for inner-shell ionization by a direct electron impact for the K-line, $Y_f(Z)$ is the fluorescence yield, $n_a$ is the density of the material atoms and $L(\omega_c^-)$ is the effective interaction length between the incident electron to the material. Typical values for the ionization cross-section of the K-shell $\sigma_K$ are $\sim 10^{-22}$ cm² for a 100keV incident electron beam. Eq. (22) captures the total number of characteristic X-ray photons emitted in all directions, yet the characteristic radiation is isotropic. Therefore, the X-ray flux collected by a detector with angular aperture $\theta_D$ and electron source current $I$ is given by

$$\dot{N}_{\text{chr}} = N_{\text{chr}}^{(t)} \theta_D^2 I/e. \tag{23}$$

### 6.2.2. Characteristic radiation from different materials

Table 5 shows the characteristic line emission for different materials, separated into rotating-anode and liquid-jet X-ray tubes. In conventional solid anode technology, the surface temperature of the anode must be below the melting point to avoid damage. To cope with the thermal load, an X-ray source based on a liquid-jet anode can be used [65–67]. Since the target material is already molten, the requirement for maintaining the target below the melting point is not essential. The current densities achievable by the liquid-jet anode are higher by two orders of magnitude than in a standard X-ray tube. However, the liquid-jet X-ray tube's average current is lower than the rotating-anode X-ray tube. Therefore, rotating-anode and liquid-jet X-ray tubes have separate purposes: the liquid-jet anode is optimized for the X-ray source brightness, whereas the rotating anode is optimized for the X-ray source flux.

**Table 5:** Characteristic X-ray parameters for copper, Molybdenum and Tungsten. The $\sigma_K$ values are valid for a 100keV electron beam energy.

| Material | K-edge energy (keV) | $L(\omega_c^-)$ (μm) | Fluorescence yield $Y_f(Z)$ | $\sigma_K$ ($\frac{barns}{atom}$) | $n_a$ ($\frac{1}{Å^3}$) | Characteristic line brightness ($\frac{photons}{s\ mm^2 mrad^2}$) |
|---|---|---|---|---|---|---|
| **Copper** (Liquid-jet anode) | 8.979 | 3 | 0.45 | 300 | 0.085 | $3.4 \times 10^{10}$ |
| **Gallium** (Liquid-jet anode) | 10.36 | 6.6 | 0.52 | 300 | 0.045 | $4.6 \times 10^{10}$ |
| **Molybdenum** (Rotating-anode) | 19.99 | 7.7 | 0.758 | 100 | 0.064 | $9.5 \times 10^8$ |
| **Tungsten** (Rotating-anode) | 69.52 | 20.7 | 0.953 | 50 | 0.063 | $1.5 \times 10^9$ |

### 6.2.3. The limitations of characteristic radiation

Characteristic radiation is the simplest operational source among the three machines since it requires a low electron energy beam (<100 keV) without the necessity for any complicated calibration processes. Moreover, its dimensions are the smallest, and the required safety shielding is the least strict due to the low electron acceleration energies (<100 keV). However, the main disadvantages of the characteristic radiation source are the lack of energy tunability and its isotropic emission. The inner shell energies of the target material anode define the emitted X-ray energies. Therefore, the X-ray application defines the anode's material as a function of the desired X-ray energy. For example, copper (~8 keV), molybdenum (~20 keV), and tungsten (~69 keV) are used for X-ray crystallography, mammography, and CT and dental imaging, respectively. This limitation restricts the use of characteristic radiation for many applications, such as K-edge absorption.

## 7. Outlook and future research directions

Developing a compact and coherent hard X-ray source has been a long-standing challenge and a major research focus in modern physics. This review has examined some of the most promising approaches for achieving this goal, with an emphasis on sources based on the coherent interaction between free electrons and matter, particularly the parametric X-ray radiation (PXR) mechanism. Recent years have seen significant progress in this field, indicating the potential for transforming these mechanisms into viable practical sources. Notably, the coherent interaction between free electrons and engineered materials emerged as a promising method for generating tunable, focused X-ray radiation without the need for large-scale facilities. The PXR mechanism, which arises from the coherent excitation of free electrons traversing periodic crystal structures, offers superior spatial coherence, high intensity in a narrow direction, and narrow bandwidth, compared to the state-of-the-art X-ray tubes. Recent advances in material engineering techniques have enabled the precise tuning of crystal structures at the atomic scale, allowing for further customization and optimization of the emitted X-ray properties.

Indeed, PXR is a prospective source of quasi-coherent hard X-rays obtainable using relatively modest electron acceleration. Although the PXR source brightness is not as high as X-rays in large facilities (**Figure 1**), it has many practical advantages: (1) Its relatively large field of view allows a short distance between the PXR source and the target, facilitating a more compact imaging environment. (2) It was demonstrated in practical applications, such as K-Edge imaging, phase-contrast imaging using differential-enhanced imaging (DEI), and computed tomography (CT). (3) Its energy is tunable using crystal rotation, permitting much flexibility in choosing the required X-ray energy. Overall, the PXR source can serve for biomedical imaging with a quasi-monochromatic and directional beam, reducing radiation dose while improving contrast.

*Prospects for phase-contrast imaging using PXR*

Future work should include research on additional characteristics of the PXR mechanism that are especially important for medical imaging applications. Experiments with lower electron source energies (<10-15 MeV) should be conducted. So far, PXR experiments for imaging applications have shown promising results using electron energies above 50 MeV [70,136,139,141]. However, a PXR source with electron energies below the neutron production threshold has many advantages, mainly less strict radiation shielding requirements and the greater availability of compact electron sources. For example, a medical linear accelerator uses a 20 MeV electron beam for radiotherapy [265]. The main challenges to overcome for lowering the electron beam energy are the higher electron scattering, the X-ray angular divergence, and the X-ray beam linewidth. Yet, these challenges are less severe for lower X-ray energy applications (e.g., mammography). Another approach for radiation reduction is to use an energy-recovery system [137].

Other research directions could use a PXR source with higher average electron beam currents to obtain in-vivo imaging. This experiment can also include a study of the PXR beam quality while moving the PXR crystal to help mitigate the electron-induced heat load. While this scheme has the potential to significantly increase the usable PXR flux, it may involve undesired artifacts such as blurring. An additional experimental validation should include the X-ray yield gain due to the usage of advanced crystal geometries in a broad energy spectrum and for different crystal materials.

*Prospects for novel developments in the PXR mechanism*

Recently, a PXR source based on a laser-plasma electron beam source has been demonstrated [73]. Electron sources based on laser-plasma accelerators usually have high energy spread and large divergence. However, the electron beam energy spread has only a small effect on the PXR emission spectrum. This scheme permits the integration of next-generation plasma-based sources into the PXR scheme, enabling high electron beam energies with compact source dimensions.

Another interesting development would include coherent PXR emission from electrons periodically modulated into micro-bunches matched to the emission wavelength. This approach enables going beyond the traditional PXR scheme analyzed above, in which the emission intensity scales linearly with the electron bunch charge. When the electrons in the bunch emit coherently, the scaling can be quadratic with the number of electrons within a pulse [3]. Recent work proposed testing a scheme of micro-bunched electrons by an XFEL creating PXR in an extremely asymmetric diffraction configuration, for which the number of produced photons was predicted to be comparable to the XFEL emission [130]. Success in such experiments will pave the way for developing new kinds of FEL facilities that utilize coherent electron interactions with matter. Future facilities of this kind could rely on PXR with both natural atomic crystals and artificial photonic crystals, to create radiation in a wide range of wavelengths from microwave and optical to X-rays [266].

Future research can enhance the PXR yield by accumulating radiation from multiple simultaneous PXR reflections in the same direction and with the same frequency. These emission channels add up coherently even when generated from different crystallographic planes of the crystal. This kind of concept as appeared in studies of the "row effect" [162,216,223] and the "plane effect" [231]. Both effects could be particularly significant at low incident electron beam energies.

An attractive prospect of PXR is the ability to generate multiple X-ray beams. Crystals typically have several crystallographic planes oriented in different directions, resulting in the simultaneous emission of multiple PXR channels, each corresponding to a different plane [184]. PXR thus offers a unique opportunity to create a facility capable of producing several X-ray beams simultaneously. This facility can function similarly to a storage ring, where multiple X-ray beams are emitted simultaneously from different stations. However, in this case, all PXR beams originate from a single crystal. A key advantage of such a multi-PXR beam facility is the relative coherence of all the beams, as they are produced by the same electrons interacting with the same crystal structure. The relative coherence offers opportunities for interferometry experiments and for pump-probe experiments with femtosecond and even attosecond time delays, depending on the pulsed nature of the incident electron beam.

Finally, an intriguing development in the PXR mechanism that can be incorporated into future PXR sources is X-ray focusing by bent crystals [71,72,183]. Conventional optical focusing components used in the X-ray spectrum are highly lossy. In contrast, a coherent interaction between the electron and the bent crystal produces a PXR beam whose phase front is curved, causing the X-ray beam to self-focus and potentially replacing the need for additional X-ray optical components.

**Data Availability**

The data and codes that support the findings of this study are available from the corresponding author upon reasonable request.

**Supplementary Material**

Supplementary material includes full mathematical derivation of all key equations in the main text and further discussion.